\documentclass[%
superscriptaddress,
reprint,
showpacs,preprintnumbers,
amsmath,amssymb,
aps,
prl,
longbibliography
]{revtex4-1}
\usepackage{color}
\usepackage{graphicx}
\usepackage{bm}
\usepackage{mathrsfs}
\usepackage{dsfont}
\usepackage{ulem}
\usepackage[unicode=true, breaklinks=false, pdfborder={0 0 1}, backref=false, colorlinks=true, linkcolor=blue, citecolor=blue]{hyperref}
\usepackage{physics}

\makeatletter

\newcommand{\Rmnum}[1]{\expandafter\@slowromancap\romannumeral #1@}
\makeatother

\begin{document}
\title{Robust Nuclear Spin Polarization via Ground-State Level Anti-Crossing of Boron Vacancy Defects in Hexagonal Boron Nitride}

\author{Shihao Ru}
\email{S. R., Z. J., H. L. and J. K. contributed equally to this work.}
\affiliation{Division of Physics and Applied Physics, School of Physical and Mathematical Sciences, Nanyang Technological University, Singapore 637371, Singapore}

\author{Zhengzhi Jiang}
\email{S. R., Z. J., H. L. and J. K. contributed equally to this work.}
\affiliation{Department of Chemistry, National University of Singapore, Singapore 117543, Singapore}
\affiliation{Joint School of National University of Singapore and Tianjin University, International Campus of Tianjin University, Binhai New City, Fuzhou 350207, P. R. China}

\author{Haidong Liang}
\email{S. R., Z. J., H. L. and J. K. contributed equally to this work.}
\affiliation{Centre for Ion Beam Applications, Department of Physics, National University of Singapore, Singapore 117542, Singapore}

\author{Jonathan Kenny}
\email{S. R., Z. J., H. L. and J. K. contributed equally to this work.}
\affiliation{Division of Physics and Applied Physics, School of Physical and Mathematical Sciences, Nanyang Technological University, Singapore 637371, Singapore}

\author{Hongbing Cai}
\affiliation{Division of Physics and Applied Physics, School of Physical and Mathematical Sciences, Nanyang Technological University, Singapore 637371, Singapore}
\affiliation{The Photonics Institute and Centre for Disruptive Photonic Technologies, Nanyang Technological University, Singapore 637371, Singapore}

\author{Xiaodan Lyu}
\affiliation{Division of Physics and Applied Physics, School of Physical and Mathematical Sciences, Nanyang Technological University, Singapore 637371, Singapore}
\affiliation{The Photonics Institute and Centre for Disruptive Photonic Technologies, Nanyang Technological University, Singapore 637371, Singapore}

\author{Robert Cernansky}
\affiliation{Institute for Quantum Optics and  Centre for Integrated Quantum Science and technology (IQST), Ulm University, Albert-Einstein-Allee 11, 89081 Ulm, Germany}

\author{Feifei Zhou}
\affiliation{Division of Physics and Applied Physics, School of Physical and Mathematical Sciences, Nanyang Technological University, Singapore 637371, Singapore}

\author{Yuzhe Yang}
\affiliation{Division of Physics and Applied Physics, School of Physical and Mathematical Sciences, Nanyang Technological University, Singapore 637371, Singapore}

\author{Kenji Watanabe}
\affiliation{International Center for Materials Nanoarchitectonics, National Institute for Materials Science, Tsukuba 305-0044, Japan}

\author{Takashi Taniguch}
\affiliation{International Center for Materials Nanoarchitectonics, National Institute for Materials Science, Tsukuba 305-0044, Japan}

\author{Fuli Li}
\affiliation{School of Physics, Xi'an Jiaotong University, Xi'an 710049, China}

\author{Teck Seng Koh}
\affiliation{Division of Physics and Applied Physics, School of Physical and Mathematical Sciences, Nanyang Technological University, Singapore 637371, Singapore}

\author{Xiaogang Liu}
\affiliation{Department of Chemistry, National University of Singapore, Singapore 117543, Singapore}
\affiliation{Joint School of National University of Singapore and Tianjin University, International Campus of Tianjin University, Binhai New City, Fuzhou 350207, P. R. China}

\author{Fedor Jelezko}
\email{fedor.jelezko@uni-ulm.de}
\affiliation{Institute for Quantum Optics and  Centre for Integrated Quantum Science and technology (IQST), Ulm University, Albert-Einstein-Allee 11, 89081 Ulm, Germany}

\author{Andrew A. Bettiol}
\email{a.bettiol@nus.edu.sg}
\affiliation{Centre for Ion Beam Applications, Department of Physics, National University of Singapore, Singapore 117542, Singapore}

\author{Weibo Gao}
\email{wbgao@ntu.edu.sg}
\affiliation{Division of Physics and Applied Physics, School of Physical and Mathematical Sciences, Nanyang Technological University, Singapore 637371, Singapore}
\affiliation{The Photonics Institute and Centre for Disruptive Photonic Technologies, Nanyang Technological University, Singapore 637371, Singapore}
\affiliation{Centre for Quantum Technologies, National University of Singapore, Singapore 117543, Singapore}
\date{\today}

\begin{abstract}
Nuclear spin polarization plays a crucial role in quantum information processing and quantum sensing. In this work, we demonstrate a robust and efficient method for nuclear spin polarization with boron vacancy ($\mathrm{V_B^-}$) defects in hexagonal boron nitride (h-BN) using ground-state level anti-crossing (GSLAC). We show that GSLAC-assisted nuclear polarization can be achieved with significantly lower laser power than excited-state level anti-crossing, making the process experimentally more viable. Furthermore, we have demonstrated direct optical readout of nuclear spins for $\mathrm{V_B^-}$ in h-BN. Our findings suggest that GSLAC is a promising technique for the precise control and manipulation of nuclear spins in $\mathrm{V_B^-}$ defects in h-BN. 
\end{abstract}

\maketitle

Optically addressable solid-state spin defects in wide band-gap materials serve as promising artificial atoms for quantum information sciences. For instance, spin defects in diamond and silicon carbide exhibit long spin coherence times, high-fidelity spin manipulations, and entangled electron-nuclear spin pairs at room temperature \cite{ref1,ref2,ref3,ref4,ref5,ref6}. In addition, layered van der Waals materials offer alternative platforms for solid-state spin defects, with reduced dimensionality facilitating scalable two-dimensional quantum device design. Among them, hexagonal boron nitride (h-BN) has attracted significant interest due to its exceptional properties, such as wide bandgap, high thermal conductivity, and robust mechanical strength \cite{ref7}. Recently, negatively charged boron vacancy ($\mathrm{V_B^-}$) defects in h-BN have been extensively studied for their coherently manipulatable spin states and high optically detected magnetic resonance (ODMR) contrast at room temperature \cite{ref8,ref9}, critical for sensitivity in quantum sensing applications \cite{ref10,ref11,ref12,ref13,ref14,ref15}. Polarization and rapid manipulation of adjacent nitrogen nuclear spins in h-BN via excited-state level anticrossing (ESLAC) have been achieved \cite{ref16}, offering avenues to store and manipulate quantum information \cite{ref17,ref18}. 

\begin{figure*}[!htbp]
\centering
\includegraphics[width=0.8\linewidth]{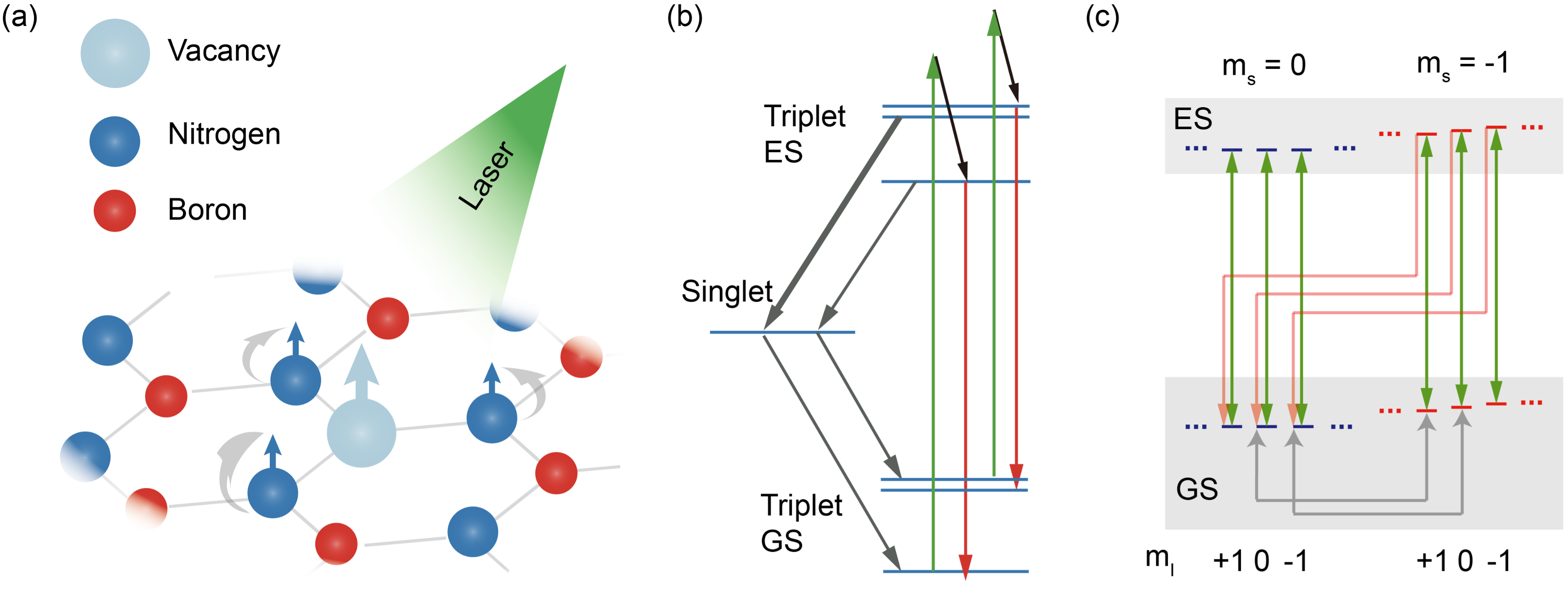}
\caption{(a) Scheme for performing nuclear polarization with the assistance of GSLAC. Magnetic field is not shown. (b) Diagram of energy levels of  $\mathrm{V_B^-}$ with optical transitions. Green, red, and grey arrows represent laser excitation, radiative recombination, and nonradiative intersystem crossing, respectively. (c) Simplified diagram of dynamics of nuclear polarization via GSLAC. Excitation and radiative decay are depicted by green arrows, while red arrows show electron spin polarization through intersystem crossing. Grey arrows illustrate flip-flop process. For simplicity, more energy levels and transitions are omitted.}\label{fig1}
\end{figure*}

The ESLAC approach to nuclear spin polarization in h-BN has limitations, such as the need for high excitation laser power and low level of nuclear spin polarization \cite{ref19,ref20,ref21,ref22}. To address the challenge, we investigate the use of ground-state level anti-crossing (GSLAC) of $\mathrm{V_B^-}$ for achieving nuclear spin polarization in h-BN. GSLAC provides several advantages over ESLAC, including increased robustness against excitation laser power fluctuations and higher nuclear spin polarization level. In this paper, we present an in-depth study of the GSLAC method and demonstrate its potential for enhancing the performance of nuclear spin polarization in $\mathrm{V_B^-}$ systems \cite{ref16,ref19,ref23}. We initiate our study by evaluating the levels of nuclear polarization at ESLAC and GSLAC regions. Our findings suggest that, even under comparatively low excitation power, GSLAC can induce high levels of nuclear polarization. Then we measure the nuclear polarization level as a function of magnetic field, aiming to find the optimal nuclear polarization level. Lastly, we have demonstrated the direct optical readout of nuclear spins through optically detected nuclear magnetic resonance (ODNMR). These comprehensive experimental endeavors pave the way for advancing quantum information processing, enhancing nuclear magnetic resonance spectroscopy sensitivity, and enabling the development of high-performance nuclear spin-based sensors.

\textit{Boron vacancy and its energy levels.} The $\mathrm{V_B^-}$ defect in h-BN is a point defect that occurs when a boron atom is missing from the crystal lattice, as shown in Fig. \ref{fig1}(a) \cite{ref8}. A $\mathrm{V_B^-}$ defect has a triplet ground state, a triplet excited state, and a non-radiative singlet metastable state. When illuminated with a green laser, $\mathrm{V_B^-}$ defects are optically excited from the ground state to the excited state. The defects can then undergo spin-dependent intersystem crossing (ISC) to the singlet metastable state, leading to spin polarization. The spin-dependent nature of the ISC processes also results in spin-dependent fluorescence, enabling optical detection and manipulation of the $\mathrm{V_B^-}$ defect's spin state. The dynamics is shown in Fig. \ref{fig1}(b) \cite{ref24}.

\textit{Nuclear polarization assisted by GSLAC.} As the external magnetic field is tuned, the energy levels of $m_s = 0$ and $m_s = -1$ states become nearly degenerate at certain magnetic field values, leading to LAC \cite{ref8,ref16,ref19,ref25}. At these points, the wavefunctions of the electronic and nuclear spin states mix, allowing for a strong interaction between them. Through this interaction, the electronic and nuclear spins can exchange spin angular momentum. This can be facilitated by a process called flip-flop, where the electronic spin flips while the nuclear spin flops simultaneously in the opposite direction. A diagram of dynamics of nuclear polarization assisted by GSLAC is shown in Fig. \ref{fig1}(c). The flip-flop process is indicated by grey arrows. As a result, the spin polarization of the electronic spin is transferred to the nuclear spins, enhancing their polarization.

\textit{Sample preparation and setup.}
h-BN flakes with thickness of several $\mathrm{\mu}$m were proton-bombarded at 250 keV energy and 3e16 $\mathrm{cm}^{-2}$ dose to create dense $\mathrm{V_B^-}$ defects \cite{ref26}.
A homemade confocal microscope were used in the experiment. A 40x/0.6 visible objective focused the excitation laser onto the sample and collected fluorescence. A 532 nm laser modulated by an Acoustic Optical Modulator provided off-resonance excitation. The photoluminescence signal, filtered through a 700-nm long pass and 1000-nm short pass filter, was coupled to a photodiode. The microwave generated by an RF signal generator and gated by an RF switch, was amplified, and applied to the sample via a coplanar waveguide where the h-BN flake was placed. A permanent magnet on a 3-axis motorized translation stage provided the magnetic field. Magnetic field alignment along the sample's c-axis was done before the experiment \cite{Supp_info}. 

\textit{Extraction of populations of nuclear spin states.} For $\mathrm{V_B^-}$ defects, the hyperfine interactions between the central electron spin and the three nearest nitrogen nuclear spins cause each electronic state to split into 7 sublevels.
This results in an ODMR peak that consists of 7 subpeaks, separated by a hyperfine value of approximately 47 MHz \cite{ref8,ref16}. 
To analyze the population of the nuclear spin states, the ODMR results are fitted with 7 peaks corresponding to these sublevels. By fitting the ODMR results with 7 subpeaks, we can extract the area of each subpeak, which is proportional to the population of the corresponding nuclear spin state \cite{ref16,ref19}. This allows us to determine the nuclear spin polarization in the $\mathrm{V_B^-}$ system and further analyze the effectiveness of the GSLAC and ESLAC techniques for achieving nuclear spin polarization.

\begin{figure}[h]
\centering
\includegraphics[width=\linewidth]{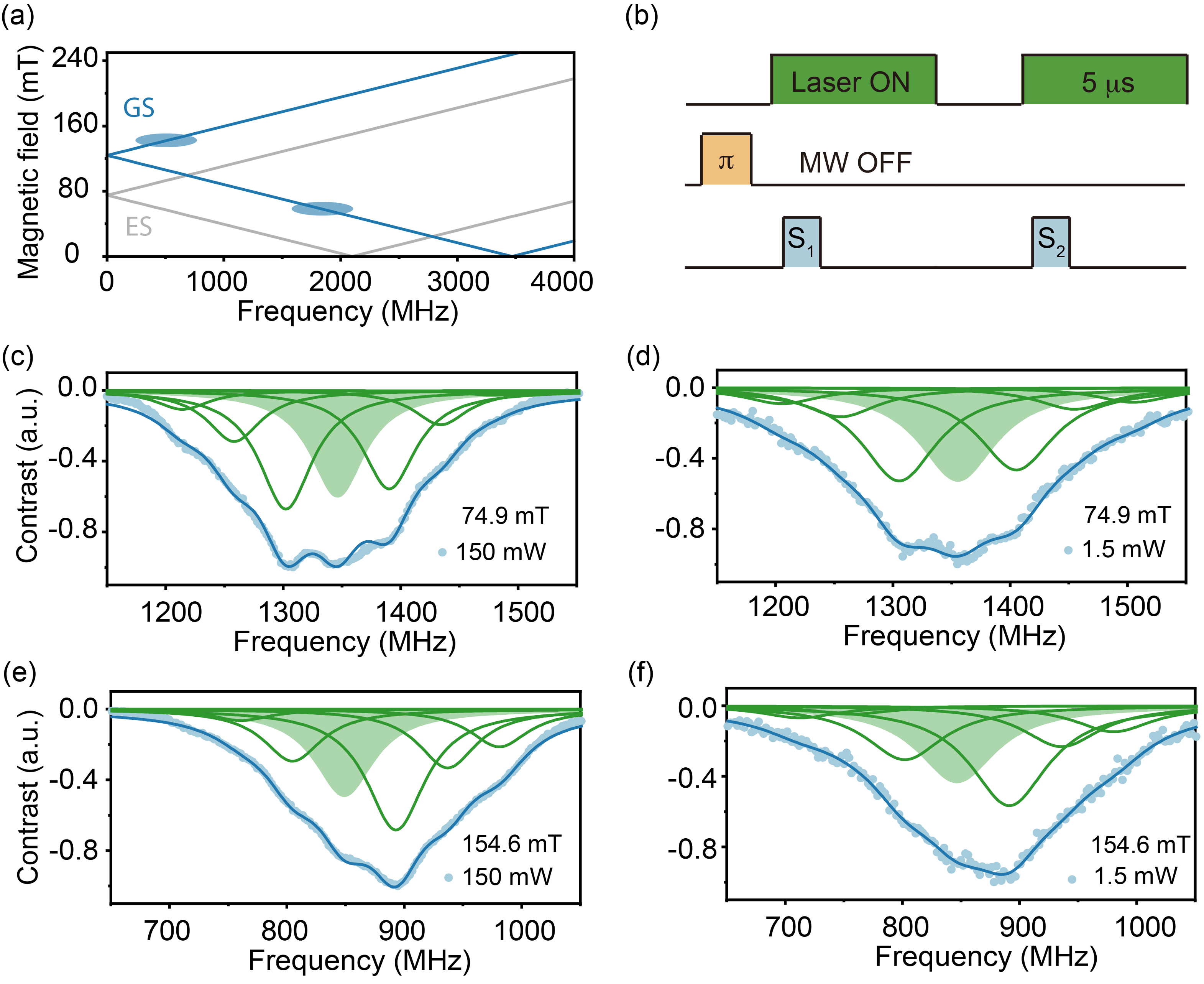}
\caption{((a) A diagram depicting the ODMR of $\mathrm{V_B^-}$ defects under an external magnetic field aligned with the c-axis. The bottom and top shaded regions correspond to the magnetic fields of (c, d) and (e, f), respectively. (b) The pulse sequence used for pulsed ODMR measurement, featuring crucial parameters labeled accordingly. Laser, microwave, and readout window are arranged from top to bottom, respectively. (c, d) Pulsed ODMR spectra captured at the magnetic field indicated by the bottom shaded area in (a), with laser powers of 150 mW (c) and 1.5 mW (d). The ODMR of $m_I$=0 is highlighted by green shaded areas for easy reference. (e, f) Pulsed ODMR spectra captured at the magnetic field indicated by the top shaded area in (a), replicating the conditions in (c, d).}\label{fig2}
\end{figure}

\textit{Laser Power Dependence in ESLAC and GSLAC: A Comparative Study.} The laser power requirements for the ESLAC and GSLAC methods in achieving nuclear spin polarization can be studied with pulsed ODMR spectra. Pulsed ODMR measurements are conducted at magnetic fields indicated by the shaded areas in Fig. \ref{fig2}(a). The lower and upper shaded fields correspond to the regions near ESLAC and GSLAC, respectively. The implemented measurement sequence is outlined in Fig. \ref{fig2}(b). We define the contrast of pulsed ODMR as $(S_1 - S_2) / (S_1 + S_2)$.

Pulsed ODMR results near the ESLAC region (74.9 mT) are analyzed with two laser powers: 150 mW (Fig. \ref{fig2}(c)) and 1.5 mW (Fig. \ref{fig2}(d)). We employ the formula,
\begin{equation}
P_{\mathrm{nuclear}}=  (\sum m_IA_I)/(3\sum A_I),
\end{equation}
to quantitatively compute the nuclear polarization levels, where $m_I$ is the quantum number of nuclear spin state and $A_I$ is the area of subpeak of $m_I$ nuclear spin state.
We observe nuclear polarization levels of $0.0655 \pm 0.008$ and $0.0343 \pm 0.024$ for 150 mW and 1.5 mW, respectively.
The relatively lower polarization at 1.5 mW suggests that ESLAC demands high laser power for increasing nuclear spin polarization, even though still in a modest level. In contrast, pulsed ODMR results near the GSLAC region (154.6 mT) are depicted in Fig. \ref{fig2}(e) and (f) for laser powers of 150 mW and 1.5 mW, respectively. Both measurements exhibit high nuclear polarization, with levels of $0.225 \pm 0.006$ and $0.170 \pm 0.023$ for 150 mW and 1.5 mW, respectively. These results underscore the efficiency of GSLAC in achieving substantial nuclear spin polarization in h-BN systems, even at lower excitation laser power, compared to ESLAC. This characteristic renders GSLAC more advantageous for practical applications where low-power operation is desired.

The robustness of GSLAC-assisted nuclear polarization is attributed to the longer residence time in the ground state prior to excitation due to light absorption.
The nuclear polarization process entails two steps: initial electron spin polarization through spin-dependent ISC under laser illumination, followed by transfer of this polarization to the nuclear spins.
The transfer occurs due to the precession of superposition states, either at the excited states (ESLAC) or the ground states (GSLAC) \cite{ref19}. A shorter residence time, such as the excited state lifetime of about 1.258 ns at room temperature (refer to \cite{Supp_info}), restricts the duration for the flip-flop process to occur. This results in the system rapidly reverting to the ground state, thus curtailing the DNP level achievable at low pump power.

\begin{figure}[!htbp]
\centering
\includegraphics[width=\linewidth]{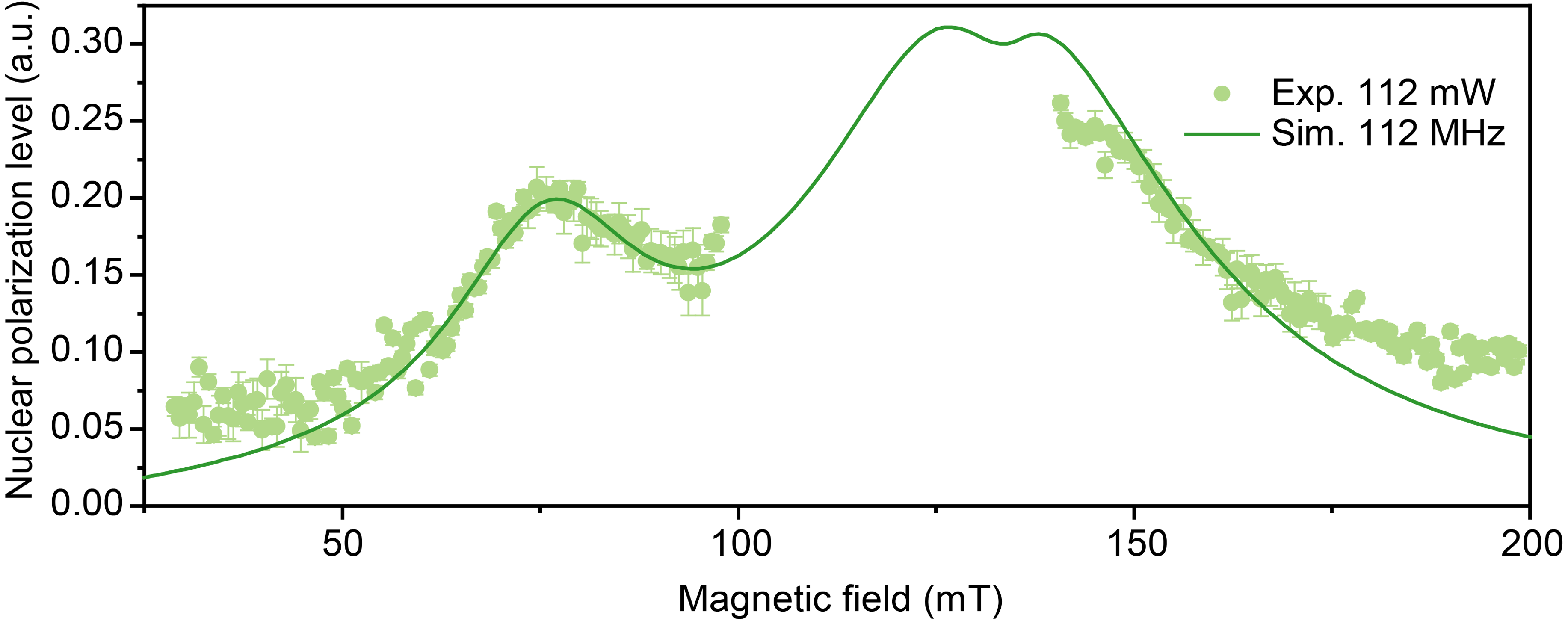}
\caption{The degree of nuclear spin polarization obtained from CW ODMR spectra is plotted against magnetic field in the range of 28 to 200 mT. The experimental results are represented by dots, while theoretical results are presented using curves. Laser power used in experiments and pump rate used in simulations have been labeled.}\label{fig3}
\end{figure}

\textit{Optimization of nuclear polarization.} To identify the magnetic field range that yields the highest nuclear polarization level, we assessed the nuclear polarization levels over a magnetic field range spanning from 25 mT to 200 mT. This span encapsulates both the ESLAC and GSLAC regions. It's important to note that the data corresponding to the precise GSLAC point, occurring around 133 mT 
in Fig. \ref{fig3}, could not be obtained from ODMR spectra due to significant spin state mixing. To accelerate the measurements and obtain reliable estimates for nuclear polarization levels, we opted for continuous wave (CW) ODMR, which provides superior signal-to-noise ratio. It should be stressed that, within the ESLAC region, CW ODMR enables greater nuclear polarization levels compared to pulsed ODMR, because of the higher average laser power.

The collected data, pertaining to a laser power of 112 mW, are depicted as green dots in Fig. \ref{fig3}. Upon the gradual increment of the magnetic field to 74.9 mT (ESLAC), we observed a peak in the nuclear polarization level. Subsequently, as the magnetic field continued to increase, the polarization level initially decreased, only to start rising again upon entering the GSLAC region. The polarization level exhibits a gradual decline beyond the GSLAC point and this trend persists with further elevation in the magnetic field. The highest nuclear polarization level observed experimentally, $0.262 \pm 0.005$, manifests around 141 mT. This magnetic field is the closest measurable point to the GSLAC in our experimental field range.

To estimate the nuclear polarization within the region near the GSLAC (100 to 140 mT), where experimental data could not be obtained, we performed a theoretical simulation. We model the system with a Hamiltonian, consisting of an electronic spin ($\mathrm{V_B^-}$) and three adjacent nitrogen nuclear spins \cite{Supp_info,ref16}. The population of each state is extracted after arbitrarily long laser interaction to ensure that a steady state is reached.
To get the levels of nuclear polarization at the strong mixing range, we extract the population of each $m_I$ state from the simulation results. The nuclear polarization levels calculated as a function of magnetic field are depicted as a green curve in Fig. \ref{fig3}. In our simulation, the nuclear polarization peaks at a level of 0.311, occurring near a magnetic field strength of 127 mT. The nuclear polarization levels predicted by our simulation align well with our experimental results within the measurable ranges, underscoring the reliability of our model \cite{ref27}.

It is important to note that the observed relatively low nuclear polarization level can be attributed to the unequal hyperfine interaction components, $A_{xx}$ and $A_{yy}$. These elements contribute to the non-diagonal hyperfine Hamiltonian, which promotes not only zero-quantum transitions but also two-quantum transitions in the vicinity of the LAC \cite{ref28,ref29}. Unlike in NV centers, this leads to unprotected highest quantum states in boron vacancies, $\ket{m_s=0, m_{I total}= +3}$, affecting nuclear polarization. Our simulations, depicted in FIG. S11, support this, showing higher polarization with equal hyperfine components. Future research could beneficially focus on strain engineering as a method to adjust the $A_{xx}$ and $A_{yy}$ hyperfine components, potentially leading to greater polarization levels \cite{Supp_info}.

\begin{figure}[h]
\centering
\includegraphics[width=\linewidth]{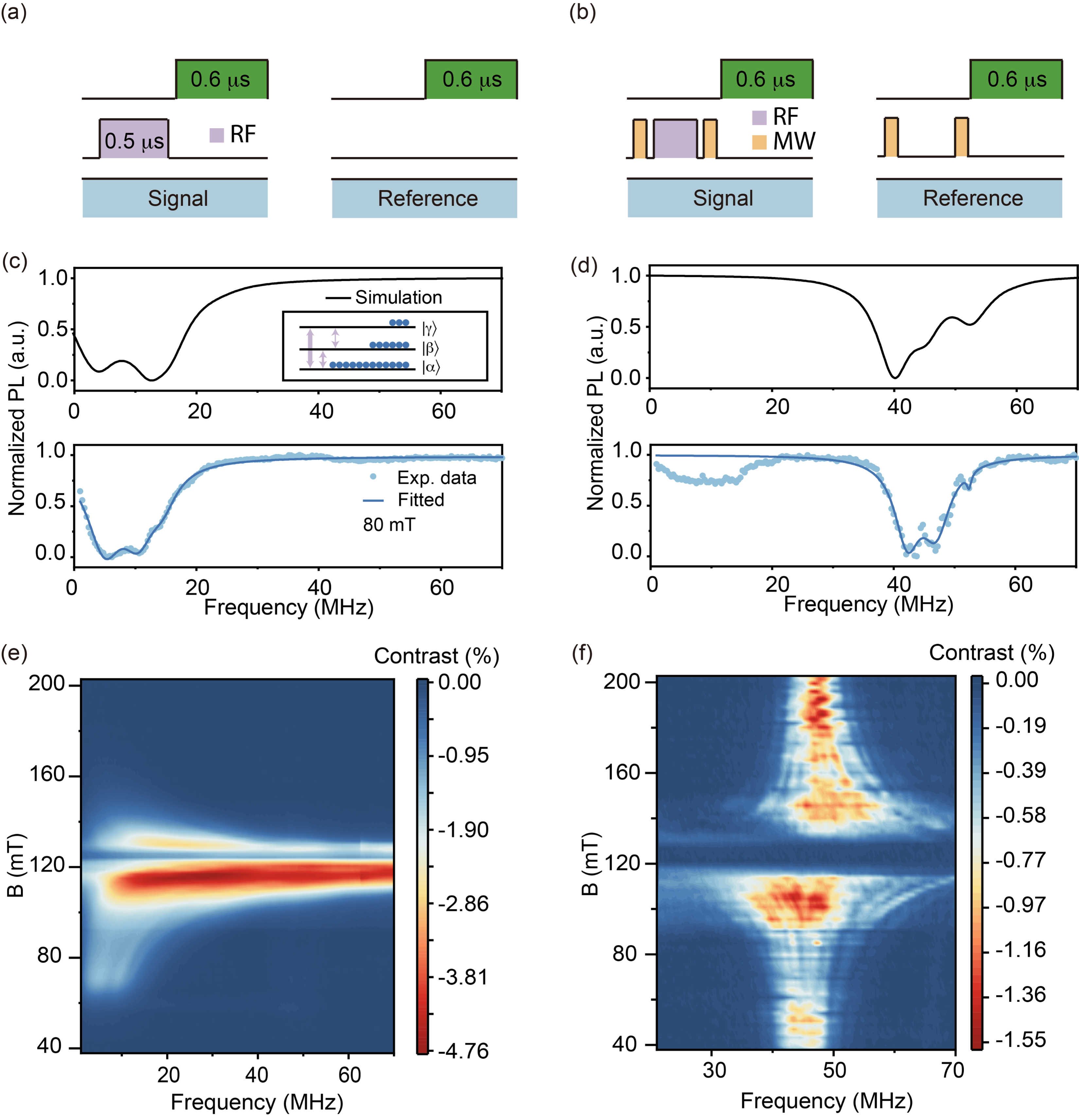}
\caption{(a), (b) Pulse sequence of ODNMR measurements for $m_s = 0$ (a) and $m_s = -1$ (b) branches. The laser, RF/MW pulses, and readout window are arranged from top to bottom. RF has a duration of 0.5 $\mathrm{\mu}$s, while each microwave has a duration corresponding to the $\pi$ pulse length. (c), (d) ODNMR spectra of $m_s = 0$ (c) and $m_s = -1$ (d) branches at 80 mT. Simulated spectra are shown in the top panels, whereas experimental results are displayed in the bottom panels. Experimental data have been fitted using multiple Lorentz peaks. Inset shows a simplified diagram of the transitions. (e), (f) Magnetic field dependent ODNMR spectra of $ m_s = 0$ (e) and $m_s = -1$ (f) branches. Due to our diplexer's limited microwave frequency range, the 113 mT to 134 mT range could not be reached in (f).}\label{fig4}
\end{figure}

\textit{Optically detected nuclear magnetic resonance.} 
The nuclear polarization induced by flip-flop can lead to the direct optical readout of nuclear spins. When only zero-quantum transitions are considered,
$\left|m_s= 0,m_{I_1,I_2,I_3}= +1,+1,+1\right>$ gives the largest fluorescence signal in general because it didn't go through the flip-flop transition.
(any flip of nuclear spins will lead to $\left|m_s= +1\right>$,
which has different energy level with $\left|m_s= 0\right>$, and therefor no mixing is happening).
Other states, for instance, $\left|m_s= 0,m_{I_1,I_2,I_3}= 0,+1,+1\right>$ can mix with $\left|m_s= -1,m_{I_1,I_2,I_3}= +1,+1,+1\right>$, which subsequently transits to a dark state via ISC.
Therefore, this reduces the fluorescence from such states since it goes through the dark state. This mechanism will allow us to read the nuclear state directly with sequence as shown in Fig. \ref{fig4}(a) and (b), an approach as ODNMR. The role of two-quantum transitions on ODNMR is discussed in the Ref. \cite{Supp_info}.

Fig. \ref{fig4}(c) shows an example of ODNMR spectrum for the transitions within spin sublevels for $\left|m_s= 0\right>$ at magnetic field of 80 mT.
The RF applied will address the transition between two states and therefore reduce the fluorescence if there is a discrepancy in the brightness of the two states, as shown in inset.
This leads to a negative value in the ODNMR contrast, which is calculated as (Signal $-$ Reference) / (Signal + Reference).
Top panel in Fig. \ref{fig4}(c) denotes the transitions as simulated. The simulation result agrees well with the experimental ODNMR spectrum shown in the bottom panel.
Similarly, if we apply additional two microwave $\pi$ pulses, this will allow us to measure the transitions between nuclear spin sublevels for $\left|m_s= -1\right>$. Fig. \ref{fig4}(d) shows an example for the transitions for spin sublevels of $\left|m_s= -1\right>$. The ODNMR signals below 30 MHz in Fig. \ref{fig4}(d) are attributable to the imperfection of the MW $\pi$ pulse, which leaves a residual population in the $\left|m_s= 0\right>$ state.

The full scan of ODNMR spectrum as a function of magnetic field is shown in Fig. \ref{fig4}(e) and Fig. \ref{fig4}(f). From Fig. \ref{fig4}(e), in the range of 100 to 140 mT, one can see that a larger ODNMR contrast can be achieved when we are approaching the GSLAC point at 128 mT and the linewidth of ODNMR transitions gets broader due to stronger mixing. At exactly 128 mT, we observe a sudden drop in the contrast of the ODNMR spectra.
This decrease can be attributed to extensive state mixing between $m_s=0$ and $m_s=-1$, resulting in reduced electron spin polarization \cite{ref27}. This is corroborated by our simulation results showcased in FIG. S12. Fig. \ref{fig4}(d) showcases the ODNMR results for the $m_s = -1$ branch. Similar conclusions to those previously discussed can be drawn from this data: as the magnetic field enters the range of 100 to 113 mT, and 134 mT to 140 mT, the transitions noticeably broaden, signifying a high degree of state mixing.

\textit{Conclusion.} In summary, we have demonstrated a robust approach for polarizing nuclear spins in h-BN, leveraging the Ground-State Level Anticrossing of $\mathrm{V_B^-}$. Notably, this method can be executed under low-power excitation. The durability of the GSLAC-assisted nuclear polarization is attributable to the extended residence time of ground states in comparison to excited states. Our simulations suggest that an optimized level of nuclear polarization can be attained near a magnetic field strength of 133 mT, where strong state mixing occurs.
We emphasize that this GSLAC-assisted nuclear polarization method is broadly applicable and not confined to the specific color center examined in this study.
Combined with dynamic decoupling methods to extend the coherence time \cite{ref_add} {and the measured coherent rotation of coupled nitrogen nuclear spins results at GSLAC \cite{Supp_info}}, the reliable polarization of nuclear spins in van der Waals materials has promising implications for quantum sensing technologies and quantum information science, including potential applications in hyperpolarization of samples \cite{ref30} and quantum registers \cite{ref31}.

\begin{acknowledgments}
S. R., J. K., H. B., X. D., F. Z., Y. Z. and W. G. acknowledge Singapore National Research foundation through QEP Grants (NRF2021-QEP2-01-P01, NRF2021-QEP2-01-P02, NRF2021-QEP2-03-P01, NRF2021-QEP2-03-P10, NRF2021-QEP2-03-P11), ASTAR IRG (M21K2c0116) and Singapore Ministry of Education (MOE2016-T3-1-006 (S)), the Australian Research council (via CE200100010), and the Asian Office of Aerospace Research and Development Grant FA2386-17-1-4064, Office of Naval Research Global (N62909-22-1-2028).
F. J. acknowledges the support of Federal Ministry of Education and Research BMBF, ERC (Synergy Grant HyperQ), European Commission (Projects FLORIN, QCIRCLE QuMICRO), DFG (Excellence Cluster POLiS,  CRC 1279, and projects 499424854, 387073854) and Carl Zeiss Stiftung.
H. L. and A. A. B. acknowledge Singapore Ministry of Education (MOE-T2EP50221-0009).
\end{acknowledgments}

\end{document}


\title{Supplementary Material for \vspace{0.08cm}\\
		\small ``Robust Nuclear Spin Polarization via Ground-State Level Anti-Crossing of Boron Vacancy Defects in Hexagonal Boron Nitride''}

    \author{{Shihao Ru$^{1, *}$, Zhengzhi Jiang$^{2,3, *}$, Haidong Liang$^{4, *}$, Jonathan Kenny$^{1, *}$, Hongbing Cai$^{1, 5}$, Xiaodan Lyu$^{1, 5}$, Robert Cernansky$^{6}$, Feifei Zhou$^{1}$, Yuzhe Yang${^{1}}$,Kenji Watanabe$^{7}$, Takashi Taniguchi$^{7}$, Fuli Li$^8$, Koh Teck Seng$^1$, Xiaogang Liu$^{2,3}$, Fedor Jelezko$^{9, \dagger}$,  Andrew A. Bettiol$^{4, \dagger}$, and Weibo Gao$^{1, 5, 10, \dagger}$\vspace{0.2cm}}
    \textit{\small \newline$^1$Division of Physics and Applied Physics, School of Physical and Mathematical Sciences, Nanyang Technological University, Singapore 637371, Singapore\\
    $^2$Department of Chemistry, National University of Singapore, Singapore 117543, Singapore\\
    $^3$Joint School of National University of Singapore and Tianjin University, International Campus of Tianjin University, Binhai New City, Fuzhou 350207, P. R. China\\
    $^4$Centre for Ion Beam Applications, Department of Physics, National University of Singapore, Singapore 117542, Singapore\\
    $^5$The Photonics Institute and Centre for Disruptive Photonic Technologies, Nanyang Technological University, Singapore 637371, Singapore\\
    $^6$Institute for Quantum Optics and IQST, Ulm University, Albert-Einstein-Allee 11, D-89081 Ulm, Germany\\
    $^7$International Center for Materials Nanoarchitectonics, National Institute for Materials Science, Tsukuba 305-0044, Japan\\
    $^8$School of Physics of Xi'an Jiaotong University, Xi'an 710049, China\\
    $^9$Institute for Quantum Optics, Ulm University, Albert-Einstein-Allee 11, 89081 Ulm, Germany\\
    $^{10}$Centre for Quantum Technologies, National University of Singapore, 117543 Singapore, Singapore\\}}

\date{\today}

\maketitle
	
\renewcommand{\theequation}{S\arabic{equation}}
\renewcommand{\thefigure}{S\arabic{figure}}

In this Supplementary Information, we provide further information on:\vspace{0.1cm}\\
  \hspace*{0.6cm}Section 1:  Magnetic Field Alignment, PL Intensity and Lifetime Measurement \\
  \hspace*{0.6cm}Section 2:  Results for ODMR Mapping and Electronic Rabi Oscillations\\
  \hspace*{0.6cm}Section 3:  Numerical Model for Nuclear Spin Polarization Simulation \\
  \hspace*{0.6cm}Section 4:  Role of Two-quantum tranistions in ODNMR\\
    \hspace*{0.6cm}Section 5:  Additional details of spin-lock, CPMG ,and XY8\vspace{0.1cm}\\
  \hspace*{0.6cm}Section 6:  Measurements of Nuclear Spin Rabi Oscillations\\
  \hspace*{0.6cm}Section 7:  Comparision of Spin Echo Spectroscopy at ESLAC and GSLAC\\
  \hspace*{0.6cm}Section 8:  Supplementary Information Figures\vspace{0.1cm}
  
  All figures and equations in Supplementary Information are labeled with the prefix ``S'' to distinguish from those in the manuscript.

\section{Section I: Magnetic Field Alignment, PL Intensity and Lifetime Measurement}\label{Sec_lifetime}

\begin{figure}[htpt]
  \centering
  \includegraphics[width=0.7\linewidth]{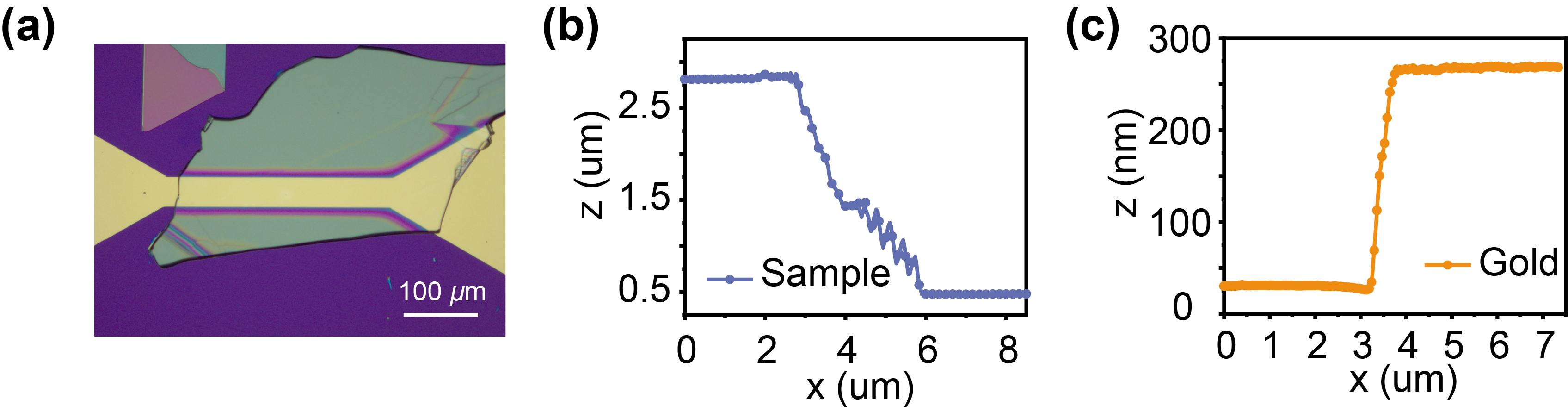}
  \caption{(a) Optical images of hBN samples. (b) AFM results of thickness of sample ($\sim 2.34$ $\mu$m) and (c) CPW ($\sim 230$ nm) measured by AFM.}\label{SI_optical_and_AFM}
\end{figure}

\noindent\textbf{Sample Preparation.}
High-pressure and high-temperature h-BN bulk crystals from the National Institute for Materials Science (Japan) are used in our work. The $\mathrm{\mu}$m-thick h-BN flakes are exfoliated from high-quality h-BN single crystal with tapes, first transferred onto a supporting block of polydimethylsiloxane on a glass microscope slide, then transferred to a Silicon wafer prepared with 200-nm thick Au-coplanar waveguide at 50 $^{\circ}$C.
The h-BN flake is irradiated using 250 keV protons with a dose of $3\times10^{16}$ cm$^{-2}$ (from a SingletonTM accelerator) over 4 mm $\times$ 4 mm to generate $\mathrm{V_B^-}$ defects.
The optical image of the sample, along with the Atomic Force Microscopy (AFM) data for thickness measurement, are presented in Fig. \ref{SI_optical_and_AFM}.\\

\begin{figure}[htpt]
  \centering
  \includegraphics[width=0.5\linewidth]{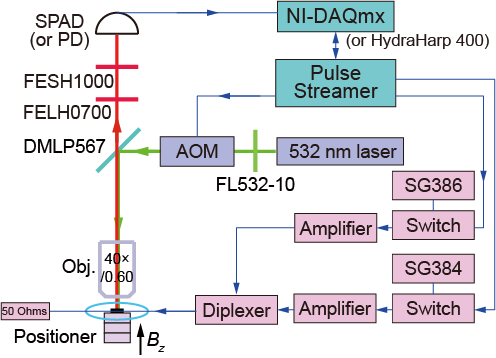}
  \caption{Optical and microwave setup for electron spin resonance measurement.}\label{Optical_Setup}
\end{figure}

\noindent\textbf{Experiment setup.} We perform all measurements with a home-built confocal microscope at room temperature (Fig. \ref{Optical_Setup}). The treated sample is mounted on an XYZ manual stage (Newport M-562F-XYZ) and an XYZ piezo scanner (Physik Instrumente P-611.3S) that is attached to a digital multi-channel piezo controller (Physik Instrumente E-727.3SD) for fine and coarse positioning $\mathrm{V_B^-}$ defects with respect to a microscope objective (Nikon S Plan Fluor ELWD 40x/0.60). A continuous-wave 532-nm laser (Sprout-H-10W DPSS) is sent through a 532-nm laser line filter (Thorlabs FL532-10), passed through an acousto-optic modulator (AOM from Gooch \& Housego 3200-121) controlled by RF Driver (Gooch \& Housego 1200AFP-AD-2.5) and directed to the objective. The filtered PL emission is separated from the excitation laser with a longpass dichroic mirror (Thorlabs DMLP567R) followed by a 700-nm longpass filter (Thorlabs FELH0700) and a 1000-nm shortpass filter (Thorlabs FESH1000). The PL emission is then focused into a multimode fiber (Thorlabs M42L02) and coupled to a photodiode (FEMTO OE-200-SI-FC or Laser Components A-CUBE-S3000-03). The counts from Photodiode are directly Analogue-to-digital converted by NI-DAQmx.

The microwave and RF signals are generated by signal generators (Stanford Research Systems SG384 and SG386), passed through microwave switches (Mini-Circuits ZASWA-2-50DRA+) and diplexer (Wainwright WDK6+5), and amplified by high-power amplifiers (Mini-Circuits ZHL-30W, ZHL-16W, ZHL-20W, and LZY-22+) with before being transmitted into the coplanar waveguide 9. All our experimental sequences, generated from a digital signal generator (Swabian Pulse Streamer 8/2) and sent to AOM driver, switches, and NI-DAQmx, are run using a custom Python data-taking software framework based on Ref. \cite{bucher2019quantum}. In addition, the lifetime measurement requires a pulse laser, single-mode fiber (Thorlabs P1-780A-FC-2), single-photon avalanche diode (Excelitas SPCM-AQRH-14-FC), and Time-Correlated Single Photon Counting (HydraHarp 400).\\

\begin{figure}[htpt]
  \centering
  \includegraphics[width=\linewidth]{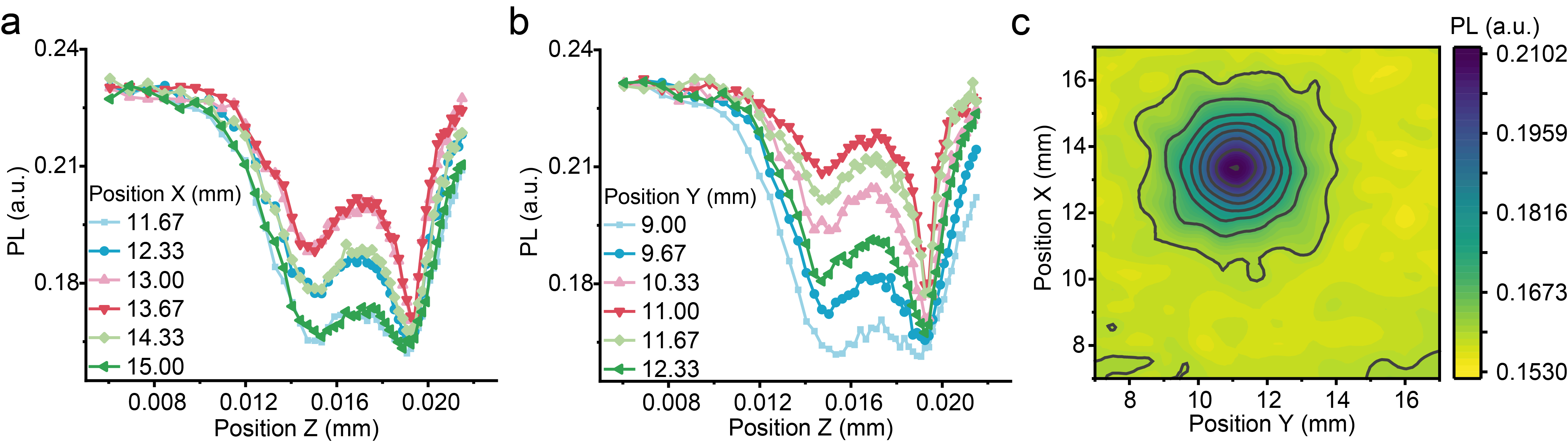}
  \caption{Alignment of magnetic field orientation. (a) First fix $y$-axis at a resonable guess value (12 mm) and scan $z$-axis PL for different $x$-axis positions. (b) Fix $x$-axis at a locally optimal location and also scan $z$-axis PL for different $y$-axis positions. (c) Fix $z$-axis at ESLAC point of $\mathrm{V_B^-}$ defects and have a PL mapping of $x$- and $y$- axis. The alignment angle of $z$-axis magnetic field (the dotted lines) is less than $0.5^{\circ}$, and is used in the subsequent experiments.}\label{S2_PL_magnetic_position}
\end{figure}

\noindent\textbf{Magnetic Field Alignment.}
The non-resonant photoluminescence (PL) signal of $\mathrm{V_B^-}$ defects exhibits a characteristic response to changes in the magnetic field, as discussed in our previous work \cite{Mu2022_prl}.
This enables us to precisely calibrate the z-axis magnetic field. The process, visualized in Fig. \ref{S2_PL_magnetic_position}, can be understood as a simple local optimization problem. The workflow comprises the following steps:

\begin{itemize}
\item[(a)] Initially, we set the y-axis to a reasonable estimated value (12 mm) and perform a z-axis PL scan for varying x-axis positions.
\item[(b)] We then set the x-axis to a locally optimal location and conduct a z-axis PL scan for different y-axis positions.
\item[(c)] Finally, we secure the z-axis at the ESLAC point of $\mathrm{V_B^-}$ defects and generate a PL map across the x- and y-axis.
\end{itemize}
{The deviation of magnetic field misalignment angle is below 0.5$^{\circ}$.}\\

\begin{figure}[htpt]
  \centering
  \includegraphics[width=0.4\linewidth]{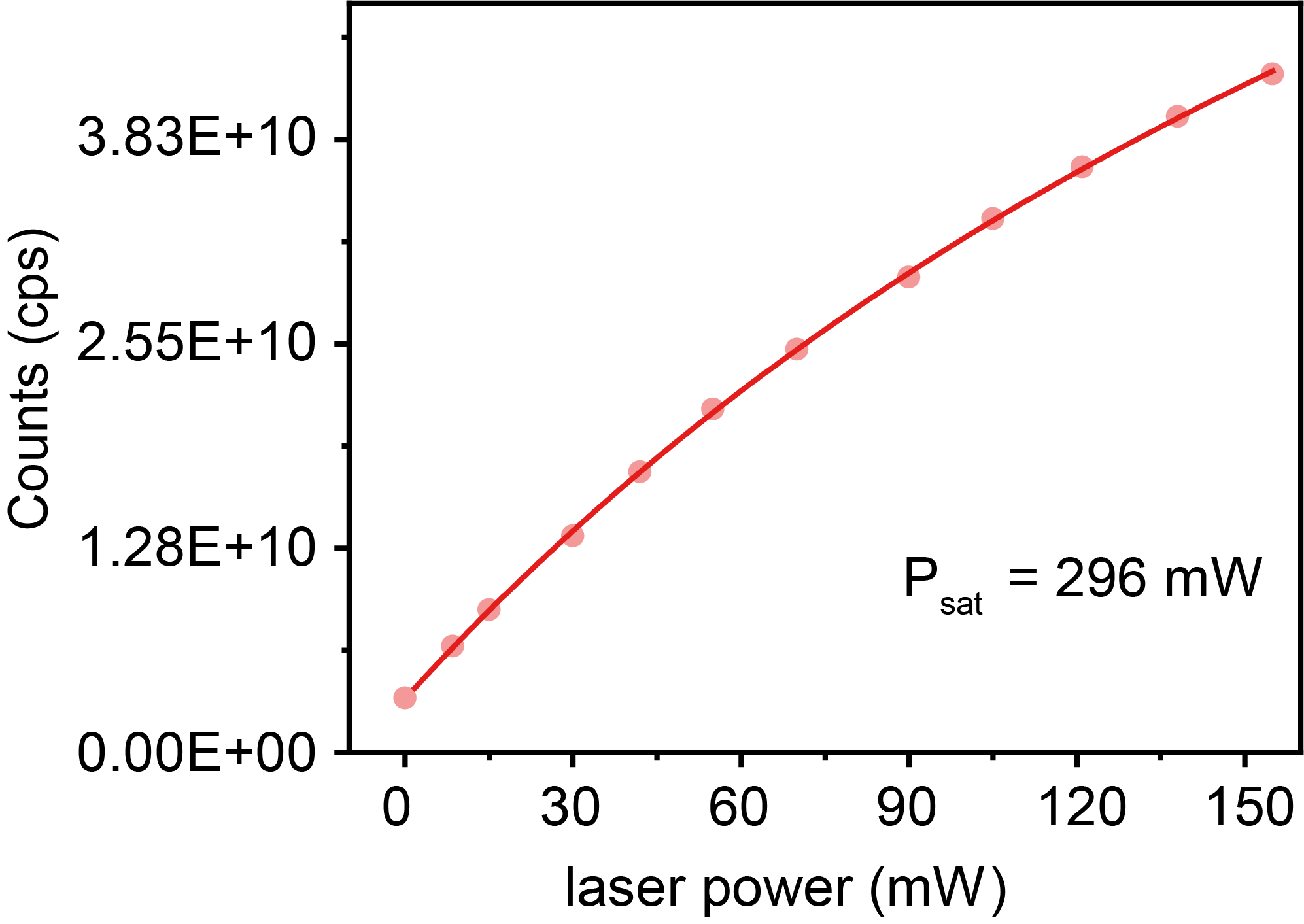}
  \caption{Count rates for the sample, subject to various laser excitation powers, are recorded. The saturation power is determined through fitting to be 296 mW.}\label{S4_PL}
\end{figure}

\noindent\textbf{PL Intensity.}
The sample we fabricated demonstrates high fluorescence intensity when illuminated with a 532 nm laser. Fig. \ref{S4_PL} presents the variation in fluorescence intensity as a function of the power of the excitation laser. These measurements are conducted utilizing a 40x objective lens with a Numerical Aperture (N.A.) of 0.6. The emitted fluorescence is subsequently directed towards an amplified photodiode (Laser Components A-Cube-S3000-03), possessing a gain of 50 MV/W at 905 nm. The voltage measurements obtained are subsequently converted into units of photon counts per second (cps).
\\

\begin{figure}[htpt]
  \centering
  \includegraphics[width=0.4\linewidth]{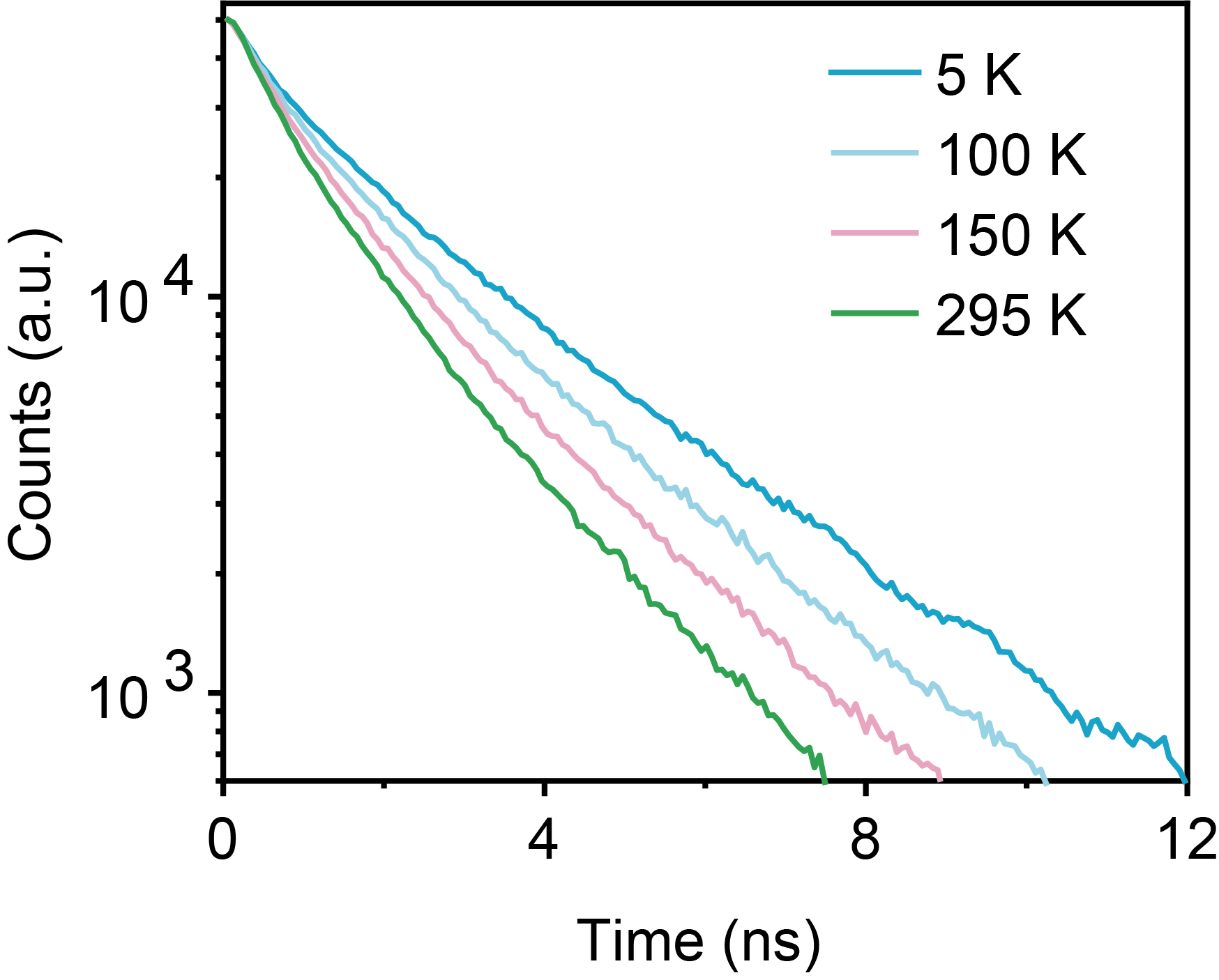}
  \caption{Lifetime measurement. (a) Temperature dependence of lifetime for Sample A. (b) Lifetime curves of Sample A and B at 5 K and 295 K.}\label{S5_lifetime}
\end{figure}

\noindent\textbf{Lifetime Measurement.}
We investigated the temperature-dependent lifetime of our sample, and the findings are illustrated in Fig. \ref{S5_lifetime}. The lifetime of the excited state is recorded as 1.29 ns at room temperature, and it extends to 2 ns at a reduced temperature of 5 K.\\

\section{Section II: Results for CW ODMR Mapping and Electronic Rabi Oscillations}

\begin{figure}[htpt]
  \centering
  \includegraphics[width=0.98\textwidth]{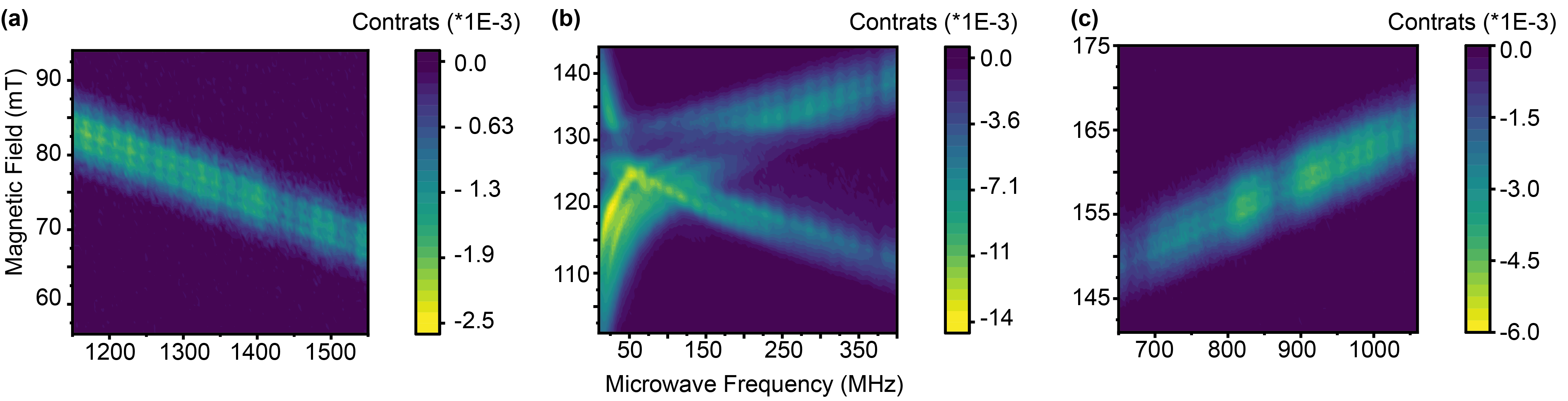}
  \caption{{Pulsed ODMR spectra at magnetic field ranging from (a) 60 – 95 mT, (b) 100 – 145 mT, and (c) 140 – 175 mT.}}
  \label{S6:full_odmr}    
\end{figure}

In the following section, we present the results of our Continuous Wave (CW) ODMR measurements conducted near ESLAC (refer to Fig. \ref{S6:full_odmr}(a)) and GSLAC (see Fig. \ref{S6:full_odmr}(b)). Both ODMR maps reveal several distinguishable lines, each corresponding to a subpeak of a distinct nuclear spin state. The ODMR at each magnetic field can be modeled using a seven-Lorentzian peak fit.\\

{We attribute this variation in ODMR contrast in Fig. \ref{S6:full_odmr}(b) to changes in nuclear polarization within these specific magnetic field regimes. In general, ODMR contrast is influenced by three factors: the transition probability of the ODMR transition, the population difference between the two energy states involved, and the difference in fluorescence intensity between these states. Due to the complex nature of the Vb system under study, it is challenging to provide a detailed quantitative analysis. However, qualitatively, we infer that higher ODMR contrast correlates with increased nuclear polarization. This inference is supported by similar phenomena observed in nitrogen-vacancy (NV) centers in diamond, which suggests that our observations are not unprecedented but rather indicative of an underlying spin-related effect.\cite{PhysRevResearch.2.023094}}\\

\begin{figure}[htpt]
  \centering
  \includegraphics[width=0.8\linewidth]{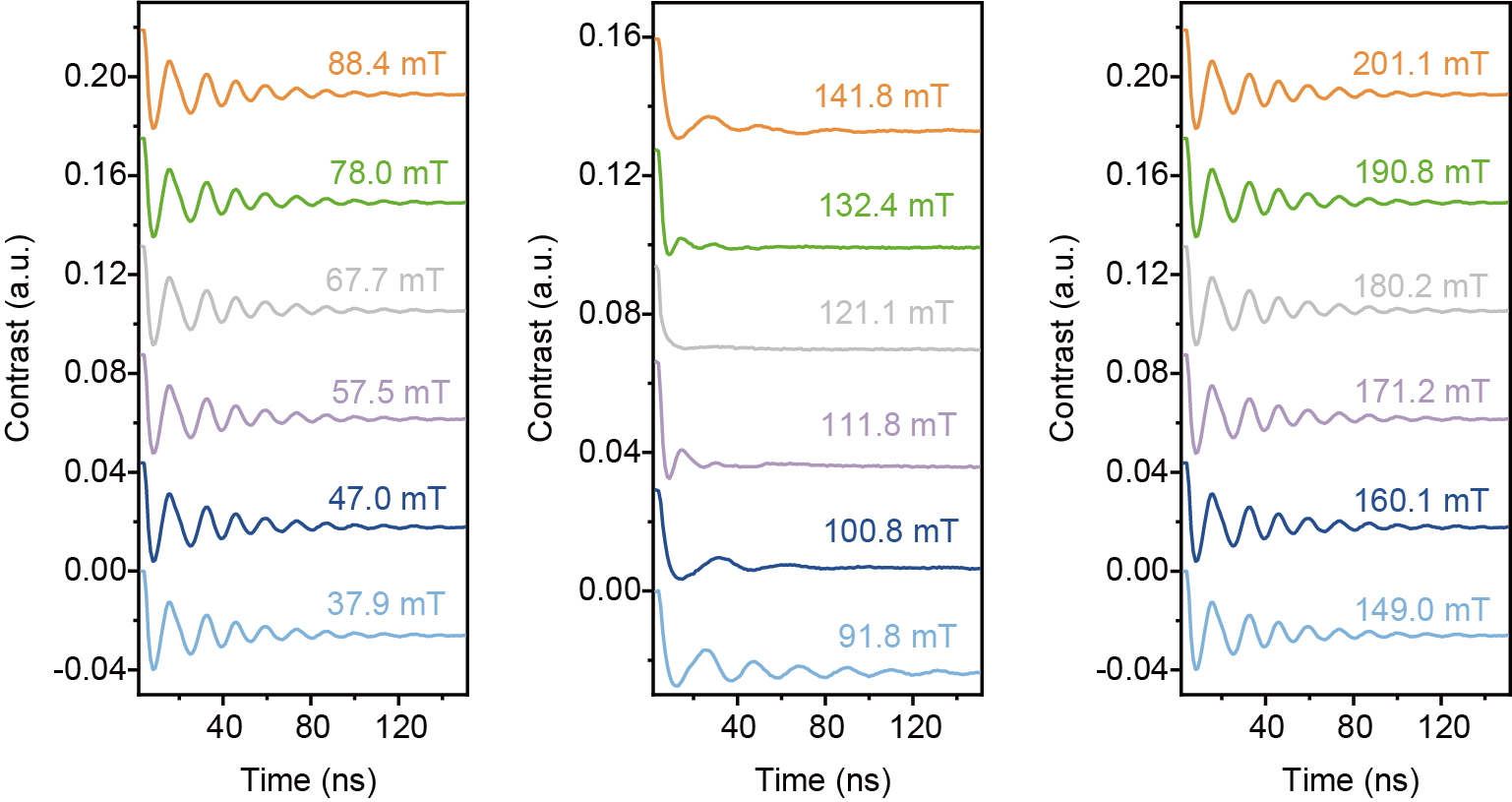}
  \caption{Electronic Spin Rabi oscillations at different magnetic fields (transition between $m_s=0$ and $m_s=-1$).}
  \label{S7: Rabi}    
\end{figure}

The findings from the Rabi oscillation experiments provide us with the $\pi$-pulse length necessary for pulsed ODMR and ODNMR measurements (Fig. \ref{S7: Rabi}). Interestingly, we observe that the Rabi oscillations exhibit modulation near GSLAC. This phenomenon arises from the pronounced state mixing between the electronic spins $m_s=0$ and $m_s=-1$, in conjunction with all their respective nuclear spin states. Moreover, it has also been reported that this observed behavior is associated with nuclear polarization \cite{liu2022coherent}.\\

\section{Section III: Numerical Model for Nuclear Spin Polarization Simulation}

\begin{figure}[htpt]
    \centering
    \includegraphics[width=0.25\textwidth]{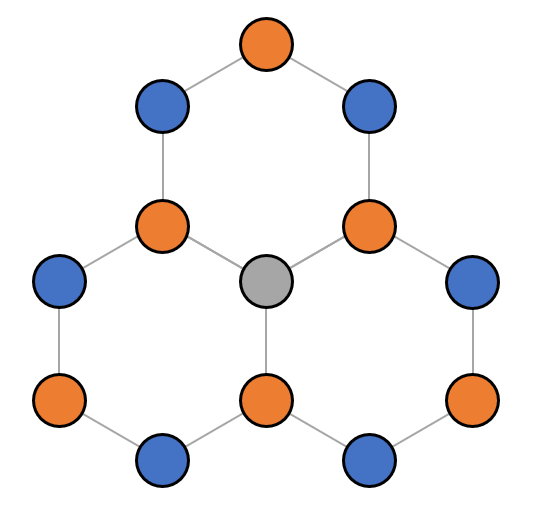}
    \caption{Simplified diagram of $\mathrm{V_B^-}$ defect (gray) surrounded by the first immediate neighbour of three ${}^{14}$N nuclear spins (orange). Blue atoms are Boron atoms.}
    \label{fig: S5_atomic_structure}    
\end{figure}

We employ the Lindblad master equation to simulate the optical cycle of the hBN system. The $\mathrm{V_B^-}$ defect is adjacent to three ${}^{14}$N nuclear spins, as shown in Figure \ref{fig: S5_atomic_structure}. As ${}^{14}$N has nuclear spin $S=1$, the hyperfine interaction between three ${}^{14}$N and the host $\mathrm{V_B^-}$ defect will cause 27 nuclear sub-levels to be observed. The Hamiltonian of hBN system is given as:
\begin{equation}
\label{eqn:simplified hamiltonian}
    \hat{H} = D \hat{S}_z^2 + \gamma_e B\cdot\hat{S}_z +\sum_{i=1,2,3}{Q\hat{I}_{z,i}^2  + \gamma_n B\cdot\hat{I}_z}  + \sum_{i=x,y,z}\sum_{j=x,y,z} \hat{S}_{i}\cdot A_{ij} \cdot \hat{I}_{j}
\end{equation}
where $\hat{S}_{x,y,z}$ and $\hat{I}_{x,y,z}$ are the $x,y,z$-projection operators of the electronic spin ($S=1$) and nuclear spin ($I=1$) respectively. $D$ is the zero-field splitting, $Q$ is the quadrupole interaction, $\gamma_{e,n}$ are the gyromagnetic ratio of the electron spin and nuclear spin respectively, and $A_{ij}$ is the $i,j$-th entry of the hyperfine interaction tensor \cite{ivady2020ab}.\\

The hBN system therefore can be described by a density operator of size $189\cross199$, in which three electron spin states with 27 nuclear spin states in the excited state (resulting in 81 states), three electron spin states with 27 nuclear spin states in the excited state (similarly, resulting in 81 states), and one electron spin shelving state with 27 nuclear spin states. We note that this shelving state is a singlet state. \\

The Lindblad master equation is given by:
\begin{equation}
\label{eqn:Lindblad}
    \frac{\partial \rho }{\partial t} = -\frac{i}{\hbar}[\hat{H},\rho]+\hat{L}(\rho)
\end{equation}
where $\hat{H}$ is the block Hamiltonian that consists of the excited, ground and the shelving state of the form:
\begin{equation}
    \hat{H} = \begin{pmatrix}
\hat{H}_e & 0_{81\cross81} & 0_{81\cross27}\\
0_{81\cross81} & \hat{H}_e & 0_{81\cross27}\\
0_{27\cross81} & 0_{27\cross81} & 0_{27\cross27}
\end{pmatrix}
\end{equation}
where $0_{m\cross n}$ is a zero matrix of dimension $m\cross n$.
The Lindblad Superoperator, $\hat{L}(\rho)$ \cite{ticozzi2009analysis} is given as:
\begin{equation*}
    \hat{L}(\rho) = \sum_k {\Gamma_k}\Big(L_k\rho L_k^{\dagger} -\frac{1}{2}\big\{L_k^{\dagger} L_k, \rho \big\}\Big)
\end{equation*}
The form of the superoperator used to model the system is the jump operator, $L_k = \ket{i}\bra{j}$, which models the non-coherent transition from state $j$ to state $i$ and $\Gamma_k$ is the associated rate of the operator. This jump operator models the non-coherent laser pump from ground to excited state ($\Gamma_p$), radiative decay from excited to ground state ($\Gamma_R$), the inter-system crossing from excited state to shelving state ($\gamma_{0,1}$), the inter-system crossing from shelving state to the ground state ($\kappa_{0,1}$) as well as the nuclear spin mixing in the shelving operators ($\Gamma_{mix}$). This mixing model is introduced phenomenologically in the shelving state manifold. The motivation is driven due to shelving state is a singlet state, while the excited and ground state are triplet states \cite{reimers2020photoluminescence}. This implies that the nuclear sublevels might not be be spin preserving when it decays through the intersystem crossing channels. This phenomenon also explains the convergence of the polarization at magnetic field away from the ESLAC and GSLAC points and stresses the importance of ESLAC and GSLAC in assisting with the nuclear polarization. Table \ref{table:param} summarizes the value of the parameters used.

\begin{table}[htpt]
\centering
\begin{tabular}{c | c} 
 \hline
 Parameter & Rate (MHz) \\ [0.5ex] 
 \hline\hline
 $\Gamma_R$ & 0.11  \cite{baber2021excited}  \\ 
 $\gamma_0$ & 220  \\
 $\gamma_1$ & 450  \\
 $\kappa_0$ & 210  \\
 $\kappa_1$ & 10  \\ 
 $\Gamma_{mix}$ & 30  \\ 
 $(A^e_x,A^e_y,A^e_z)$ & (23.06, 45.79, 23.07)  \\
 $(A^g_x,A^g_y,A^g_z)$ & (69.17, 137.36, 47.94)  \\[1ex] 
 \hline
\end{tabular}
\caption{Parameter values for the numerical calculation}
\label{table:param}
\end{table}

Due to their energy level degeneracy, their effective energy level splitting can be represented by 1 nuclear spin with spin $S=3$, with $m_I=+3,+2,+1,0,-1,-2,-3$ and the degeneracy of  respectively.

To acquire the numerical solution for the polarization, we extract the population count of the 27 nuclear sublevels within the $m_{s,g}=0$ manifold of the density operator $\rho$. Due to their degeneracy, we can represent the 27 nuclear sublevels using the Dicke state of $m_I=\pm3,\pm2,\pm1,0$ with the degeneracy ratio $1:3:6:7:6:3:1$. We can proceed to calculate the nuclear polarization using this formula:
\begin{equation}
    P(B) = \frac{\sum_{k=-3}^{3} k g_k \rho(m_s =0, m_I = k)}{\sum_{k=-3}^{3} g_k \rho(m_s =0, m_I = k)}
\end{equation}
where $g_k$ is the degeneracy of the $m_I$ states ($1:3:6:7:6:3:1$) and $\rho(m_s =0, m_I = k)$ is the population sum of the density operator entry that represents the $m_s =0, m_I = k$ state.\\

\noindent\textbf{Mixed eigenstates of $H_{\parallel}$.}
For NV centers in diamond, the states $\ket{m_s = 0, m_I = +1}$ and $\ket{m_s = -1, m_I = -1}$ are still the eigenstates of full Hamiltonian $H_{Total}$, if we only consider the transitions between $m_s = 0$ and $m_s = -1$.
However, in $V_B^-$ system, all the eigenstates of $H_{\parallel}$ are mixed because of the two-quantum transitions.
$\ket{m_s = -1, m_I = -3}$ and $\ket{m_s = 0, m_I = +3}$ are already not the eigenstates of full Hamiltonian.
When the magnetic field approaches the GSLAC, $\ket{m_s = 0, m_I = +3}$ is heavily mixed, which limits the degree of nuclear polarization.\\

\begin{figure}[htpt]
  \centering
  \includegraphics[width=0.7\textwidth]{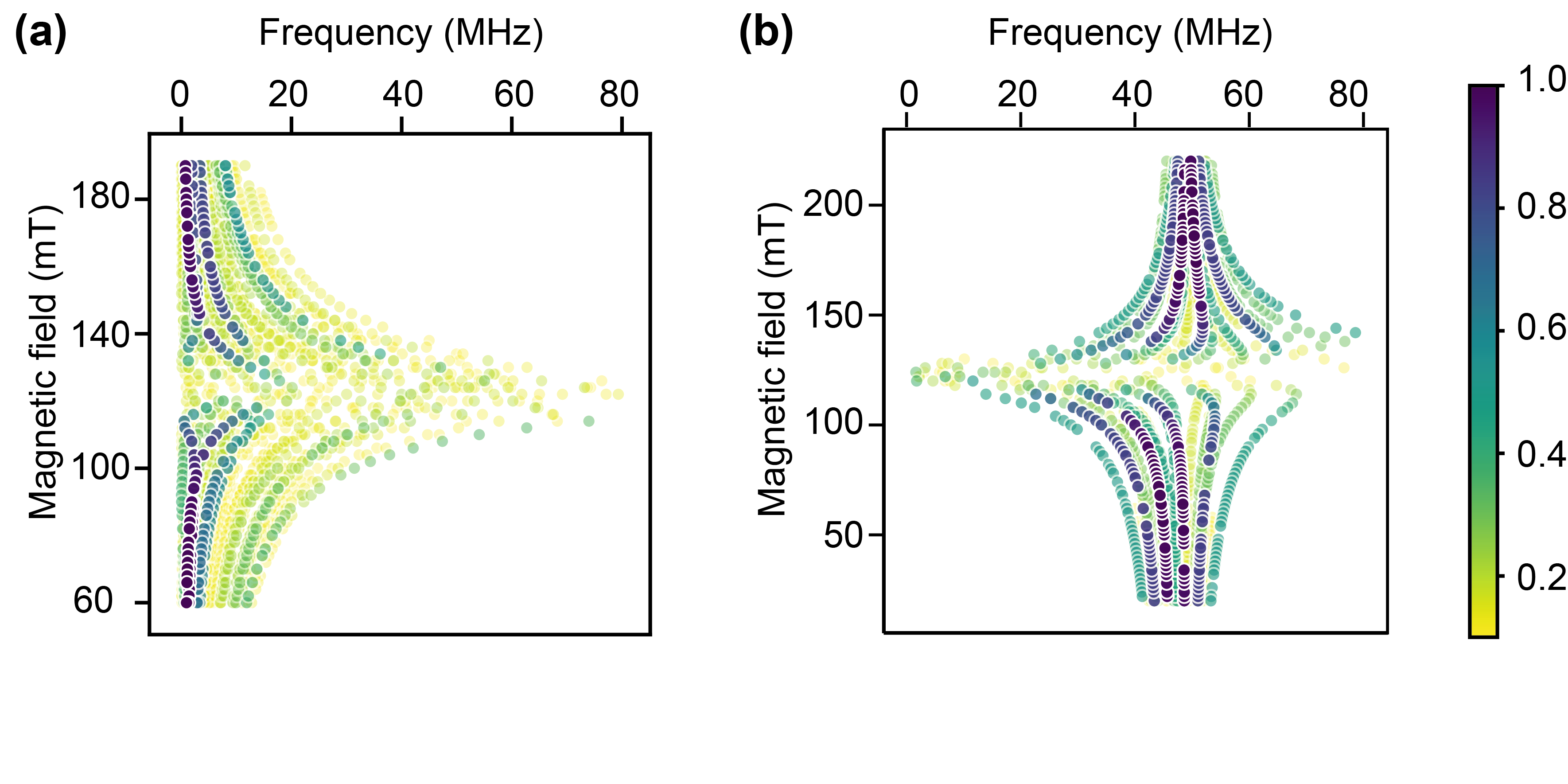}
  \caption{Simulated transtions of ODNMR results. (a) Simulated ODNMR transitions of $m_s$ = 0 branch and (b) $m_s$ = -1 branch.}
  \label{S9_ODNMR}    
\end{figure}

\noindent\textbf{Simulation of ODNMR transitions.}
Our model is also employed to simulate transitions observable in the ODNMR spectra. The outcomes of these simulations, presented in Fig. \ref{S9_ODNMR}, exhibit a good agreement with the findings derived from our experimental studies.\\

\section{Section IV: Role of Two-quantum tranisitons in ODNMR}
\noindent\textbf{Eigenstates of $H_{Total}$.}
Taking the eigenstates of $H_{\parallel}$ as the basis, each eigenstate of $H_{Total}$ can be depicted as a linear combination of those basis states whose quantum numbers ($m_s + m_I$) are either identical or differ by two. The mix of states with equivalent quantum numbers results in zero-quantum transitions, which is a phenomenon observed in the NV center.
The mix of states with quantum numbers that differ by two arises due to the different transverse hyperfine components ($A_{xx} \neq A_{yy}$) of $\mathrm{V_B^-}$. 
Considering only zero-quantum transitions, we can express an  as follows:
\begin{equation}
\ket{a^{\prime}}=a_1\ket{m_s=0,m_{I(1+2+3)}=-3}+b_3\ket{m_s=-1,m_{I(1+2+3)}=-2}
\end{equation}
however, when we include both zero- and two-quantum transitions, an eigenstate would be expressed as
\begin{align}
\ket{\alpha}&=a_1\ket{m_s=0,m_{I(1+2+3)}=-3}+b_3\ket{m_s=-1,m_{I(1+2+3)}=-2}\notag\\
&+c_6\ket{m_s=0,m_{I(1+2+3)}=-1}+d_7\ket{m_s=-1,m_{I(1+2+3)}=0}\notag\\
&+e_6\ket{m_s=0,m_{I(1+2+3)}=+1}+f_3\ket{m_s=-1,m_{I(1+2+3)}=+2}\notag\\
&+g_1\ket{m_s=0,m_{I(1+2+3)}=+3}
\end{align}
In these equations, the variables a, b, c, d, e, f, and g denote the corresponding coefficients. The subscripts of these variables indicate the associated degeneracies.\\

\noindent\textbf{Brightness of eigenstates of $H_{Total}$.}
The 'brightness' of the eigenstates of $H_{Total}$ can be determined by summing their projection onto the basis states with $\ket{m_s = 0}$, irrespective of their $m_I$ values. This is owing to the spin-dependent ISC process. An eigenstate with a more significant projection onto the basis with $\ket{m_s = 0}$ is less likely to undergo ISC and, as a result, appears as a 'brighter' state.\\

\noindent\textbf{Polarization of eigenstates of $H_{Total}$.} 
The different brightness of these eigenstates leads to a variation in their populations. “Darker” states, having a larger projection onto the basis with $\ket{m_s = -1}$, are more likely to undergo ISC. Consequently, their quantum numbers have a higher probability of increasing by one. This implies that they are more susceptible to being “ejected” from their existing eigenstates, which are composed of basis states with quantum numbers differing by zero and two. Ultimately, the “brighter” eigenstates, which exhibit a lower likelihood of such “ejection”, maintain higher populations.\\

It is worth noting that the presence of two-quantum transitions can detrimentally influence the degree of nuclear polarization. In the absence of two-quantum transitions, and only with zero-quantum transitions, the increase in quantum numbers due to ISC would be irreversible. This process would ultimately result in an accumulation of most states at the highest quantum number state. However, two-quantum transitions introduce pathways for quantum number decrement, thereby reducing the polarization level in $V_B^-$ compared to the NV center in diamond.\\

\noindent\textbf{ODNMR signal.} After considering the effect of two-quantum transitions, we still have discernable differences in both brightness and population between the eigenstates. These disparities form the basis for the visibility of the ODNMR signal. During an ODNMR measurement, a radio-frequency (RF) applied as a $\pi$ pulse results in population swapping between two states, provided their energy difference is in resonance with the applied RF. This interchange leads to a reduction in fluorescence, allowing for the detection of the ODNMR signal.

\section{Section V: Details of spin-lock, CPMG ,and XY8}
To implement dynamic decoupling for the boron vacancy, we employed the AWG70002A from Tektronix in conjunction with an RF signal generator SG384 from Stanford Research Systems.
The SG384 generates a base frequency of 1600 MHz, which is combined with a 400 MHz signal modulated by the AWG70002A using a mixer. This combined signal passes through a high pass filter, eliminating the 1200 MHz component and preserving the 2000 MHz component. The resultant high-frequency signal is then amplified by a microwave amplifier before being transmitted to the co-planar waveguide. Additionally, the AWG70002A controls the phase and amplitude of this microwave signal.
The Spin-lock, CPMG, and XY8 sequences' effectiveness in extending coherence time are evaluated in the subsequent sections.\\

\begin{figure}[htpt]
  \centering
  \includegraphics[width=0.7\textwidth]{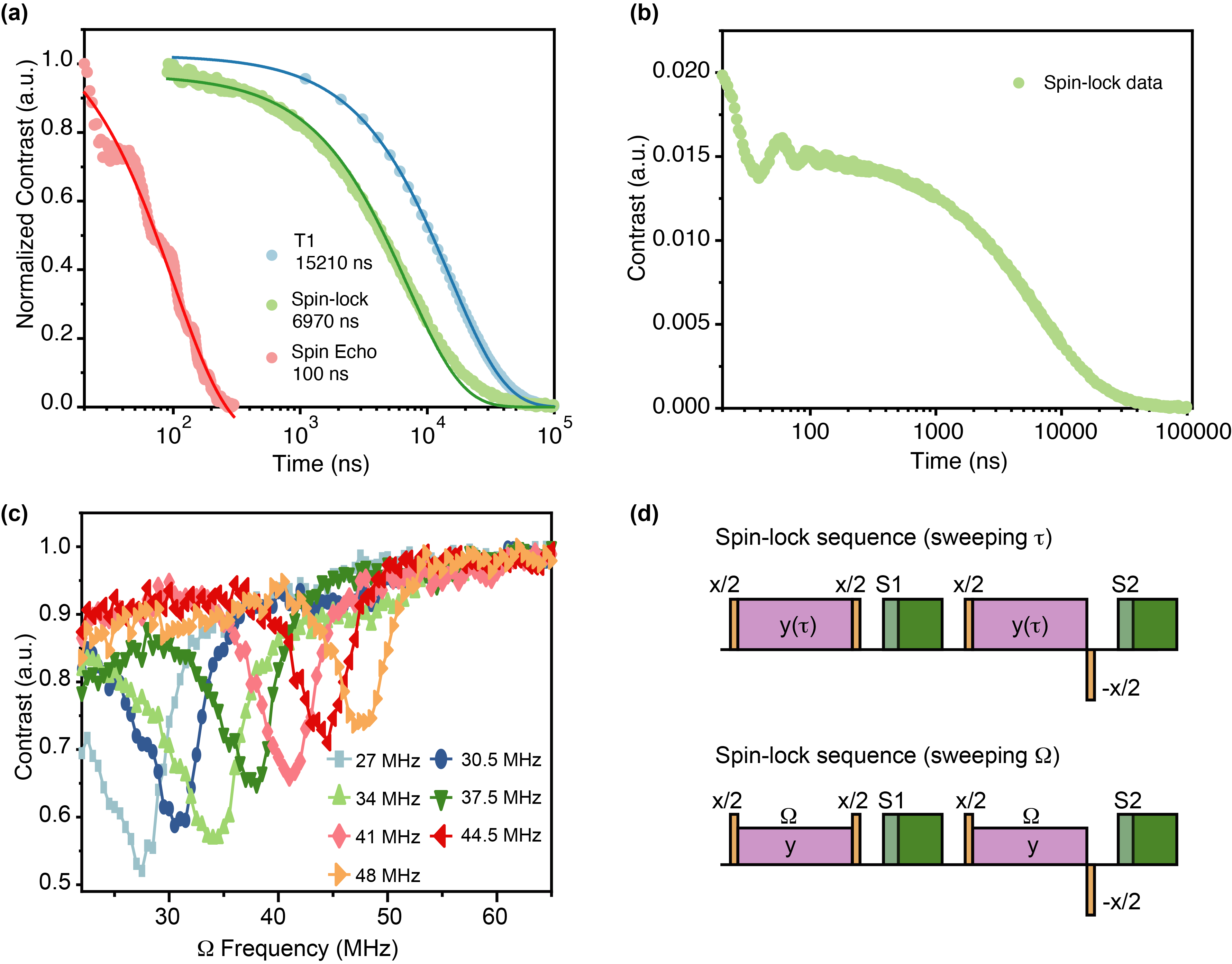}
  \caption{(a) Results of relaxation time T1 (in blue), Spin-lock (in green), and Spin echo (in red) measurements. The driving frequency for the spin-lock is set at 31.25 MHz. The measured T1 and corresponding coherence times are indicated in the labels. (b) A detailed view of the spin-lock measurement results with a 31.25 MHz driving frequency. (c) Spin-lock results obtained with applying radio-frequency (RF) signals at varying frequencies. The frequency of each RF signal is labelled accordingly. (d) The sequence of spin-lock measurements (top) and the sequence used to perform RF sensing utilizing the spin-lock technique (bottom).}
  \label{S11}    
\end{figure}

\noindent\textbf{Spin-lock.} To find the efficacy of spin-lock in enhancing coherence time, we first compared spin-lock with spin echo \cite{London_PRl}. The spin-lock sequence comprises two half-pi pulses with phases in the x-direction, separated by a y-direction pulse lasting $\tau$ us. The results, including spin-lock, spin echo, and relaxation time (T1) measurements, are presented in FIG. \ref{S11}. Notably, the coherence time was extended to 6.97 µs with spin-lock, approximately 70 times longer than the coherence time measured with spin echo (100 ns), and comparable to T1 (15.21 µs). This substantial extension in coherence time underscores the effectiveness of spin-locking in decoupling the spins from their environment. \\

We observe that the performance of spin-locking is closely linked to the microwave power used. A higher microwave power tends to yield a longer coherence time, with its theoretical limit being the relaxation time (T1). This finding aligns with the conclusions presented in \cite{PhysRevA.98.033409}. To effectively decouple the boron vacancies, we employed a spin-lock driving strength of 31.25 MHz. However, this high driving frequency inadvertently also affects the nearest ±1 hyperfine energy levels. As shown in FIG. \ref{S11} (b), noticeable oscillations are evident in the spin-lock results during the initial 150 ns. This limits further advancement of spin-lock related sequences in boron vacancies. For example, achieving nuclear spin polarization through the Hartmann-Hahn condition, as detailed in \cite{London_PRl}, is one such approach impacted by this limitation. The Hartmann-Hahn condition is highly sensitive to pulse errors and detuning. As mentioned earlier, when dealing with boron vacancies in hBN, a strong microwave centred at the 0 hyperfine state doesn't just target the intended hyperfine state, but also influences the adjacent ±1 hyperfine states. This interaction results in both pulse length errors and frequency detuning.\\

Spin-lock can find application in the measurement of radio-frequency (RF) signals. We measured RF signals with different frequencies with spin-lock technique, as illustrated in FIG. \ref{S11} (c). In this context, the spin-lock sequence involves applying an RF field to the boron vacancies and sweeping the microwave driving frequency ($\Omega$) of the spin-lock's y-pulse from 20 MHz to 65 MHz, as shown in FIG. \ref{S11} (d) bottom. This technique is based on the principle that when the driving frequency is in resonance with frequency of the applied RF field, the electron spins interact with the RF field, accumulating a phase. Consequently, this interaction alters the population of the final projected electronic spin states 0 and -1.\\

\begin{figure}[htpt]
  \centering
  \includegraphics[width=0.7\textwidth]{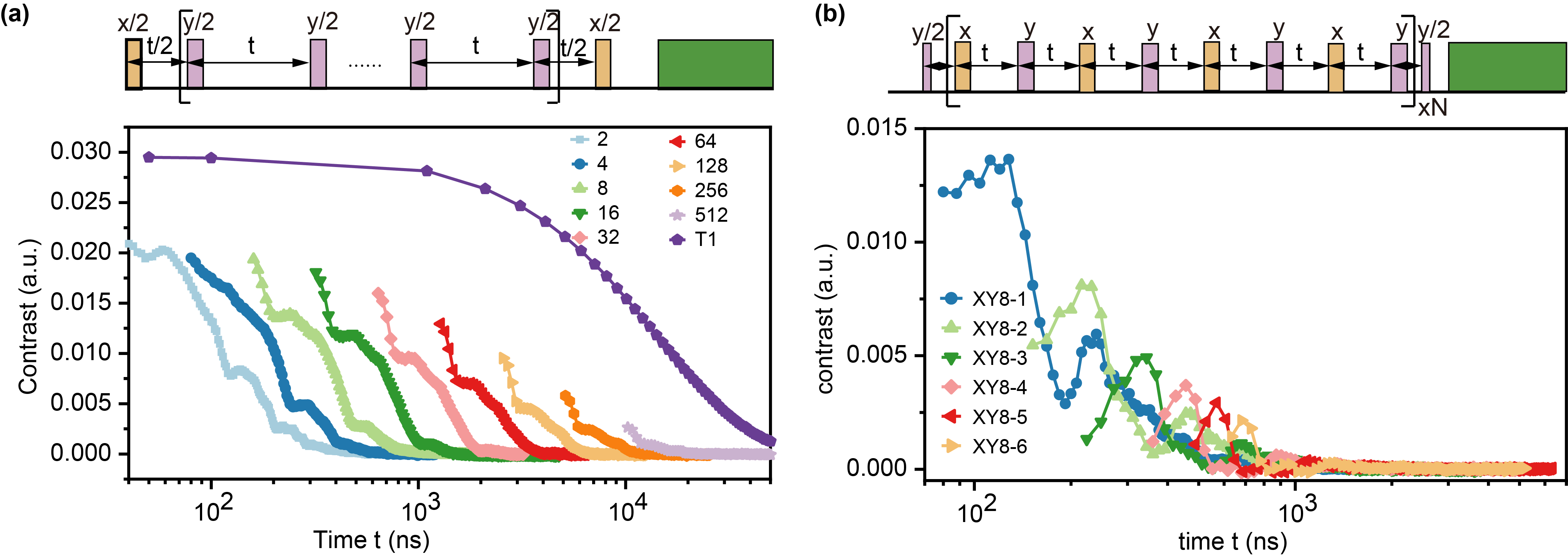}
  \caption{(a) CPMG experiment results. The number of pi pulses are indicated in label. (b) XY8-N experimental results. }
  \label{S12}    
\end{figure}

\noindent\textbf{CPMG-N and XY8-N.}
We subsequently conducted CPMG-N and XY8-N measurements to examine their influence on prolonging coherence time. FIG. \ref{S12} displays the decoherence curves for the electronic spins of boron vacancies, obtained using the CPMG-N decoupling pulse sequences, where N represents the number of pi pulses. These CPMG-N sequences are composed of an initial (pi/2)x pulse, followed by a repeating pattern of [ - $\tau$ - (pi/2)y ]*N, and concluding with a final (pi/2)x pulse for optical readout to map the accumulated phase onto spin populations. Notably, using CPMG-512, the coherence time was extended to 2.70 ± 0.41 µs, as shown by the shallow purple curve in FIG. \ref{S12} (a). For comparison, the T1 value, recorded as 15.214 ± 0.035 µs, is also provided. The results of the XY8-N measurements are presented in FIG. \ref{S12} (b). In this case, the XY8-N sequence involves 8*N pi pulses. It is evident from these results that the XY8 sequence is less effective in extending coherence time compared to the CPMG sequence. This could be possibly caused by the effect of spin locking and strong coupling with the surrounding nitrogen nuclear spins \cite{senkalla2023germanium}.

\section{Section VI: Measurements of Nuclear Spin Rabi Oscillations}

\begin{figure}[htpt]
  \centering
  \includegraphics[width=0.8\textwidth]{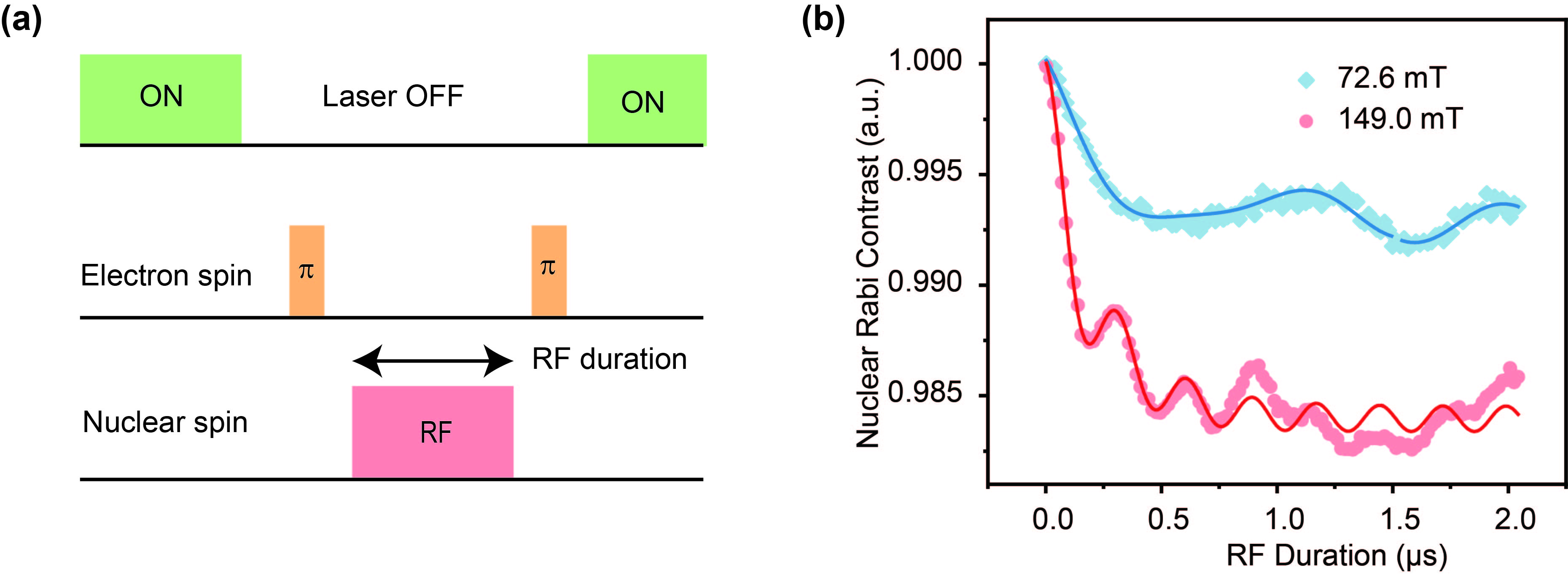}
  \caption{{(a) Sequence of nuclear spin measurement. (b) Nuclear spin rabi results at 72.6 mT (ESLAC) and 149.0 mT (GSLAC).}}
  \label{S15}    
\end{figure}

The sequence for nuclear spin Rabi oscillations is depicted in Figure \ref{S15}(a). Results from nuclear Rabi measurements at both ESLAC and GSLAC are presented in Figure \ref{S15}(b). The data were modeled using three damped sine functions. The highest frequency from these sine functions is selected to establish the nuclear Rabi frequency. Given that the RF power remains constant, an increase in the nuclear Rabi frequency indicates a significant boost in the effective gyromagnetic ratio of the nuclear spins. We observed the nuclear Rabi frequency to be 0.96 MHz near ESLAC, whereas it increased to 2.15 MHz in the vicinity of GSLAC.\\

At ESLAC, the enhancement factor for the gyromagnetic ratio of nitrogen spins, denoted as $\gamma_{N,eff} / \gamma_{N,bare}$, has previously been reported to be 350 \cite{gao2022nuclear}. As the magnetic field nears GSLAC, this enhancement factor escalates, reaching 783 at a magnetic field strength of 147mT. This increase in nuclear spin manipulation speed allows for a higher number of quantum operations within the same timeframe, offering a potential strategy to mitigate the limitations posed by the short electron coherence time T2.

\section{Section VII: Comparison of spin echo spectroscopy at ESLAC and GSLAC}

\begin{figure}[htpt]
  \centering
  \includegraphics[width=0.4\textwidth]{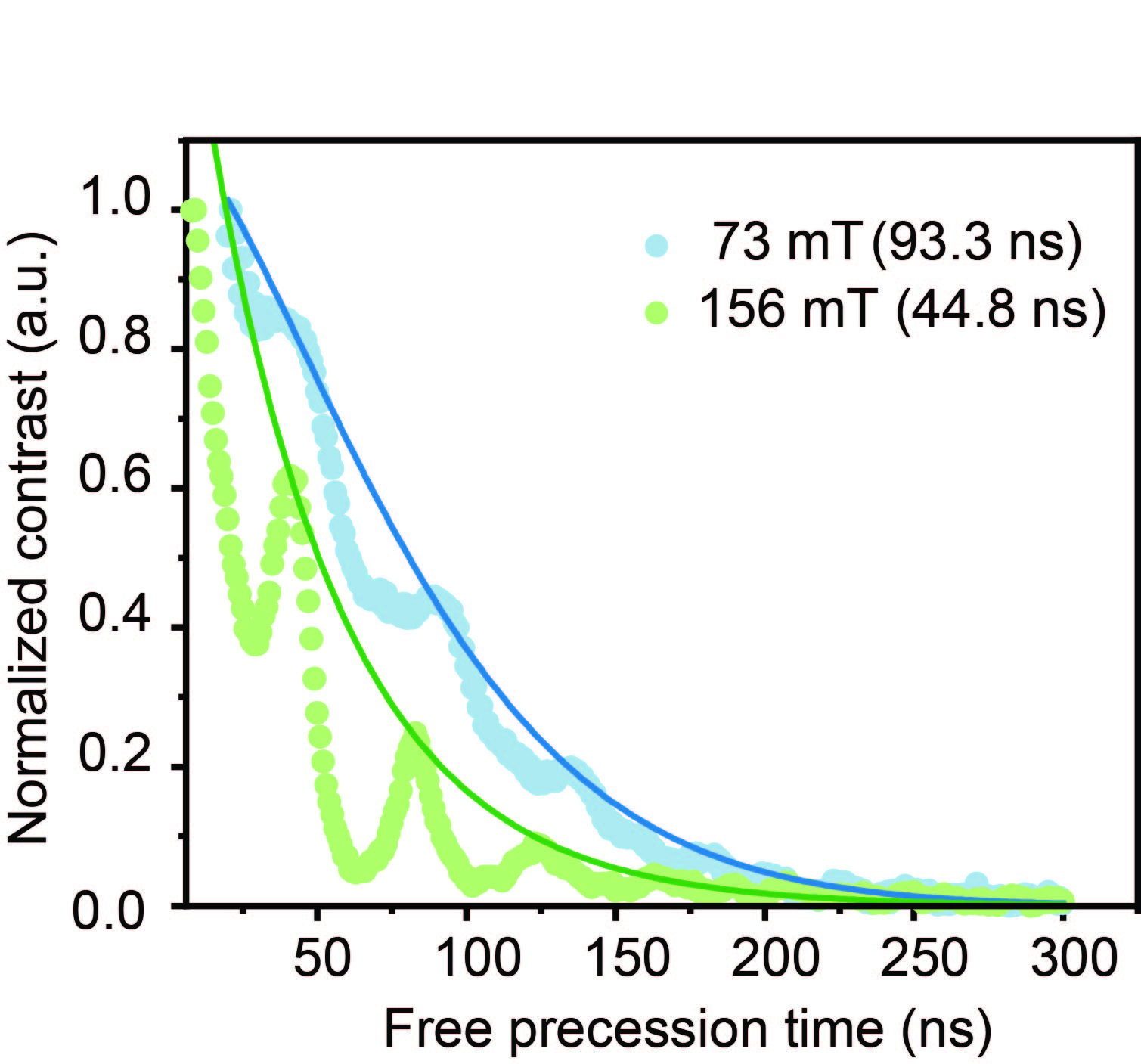}
  \caption{{Comparison of spin echo spectroscopy at magnetic field of 73 mT (ESLAC) and 156 mT (GSLAC).}}
  \label{S16}    
\end{figure}

We have compared these spin echo spectroscopy at GSLAC (156 mT) with those obtained at ESLAC (73 mT), and the results are illustrated in Fig. \ref{S16}. We noted a pronounced modulation in the spin echo spectra at GSLAC attributable to interactions with polarized nuclear spins. Specifically, a noticeable dip in the Vb fluorescence signal at an evolution time of 28 ns in the GSLAC spectra can be traced back to coherent interactions with polarized nitrogen spins. This distinct signal, weakened at ESLAC, becomes evident only under the conditions influenced by GSLAC. The appearance of the dip in free evolution spectroscopy was traditionally approached using extended dynamical decoupling sequences like XY8-N to prolong the interaction time between electron and nuclear spins \cite{PhysRevLett.109.137601}. Remarkably, in our case, this outcome was facilitated simply by adjusting the magnetic field to engage GSLAC, suggesting an intensified coupling between nuclear and electron spins under these specific conditions.\\

\begin{figure}[htpt]
  \centering
  \includegraphics[width=1\textwidth]{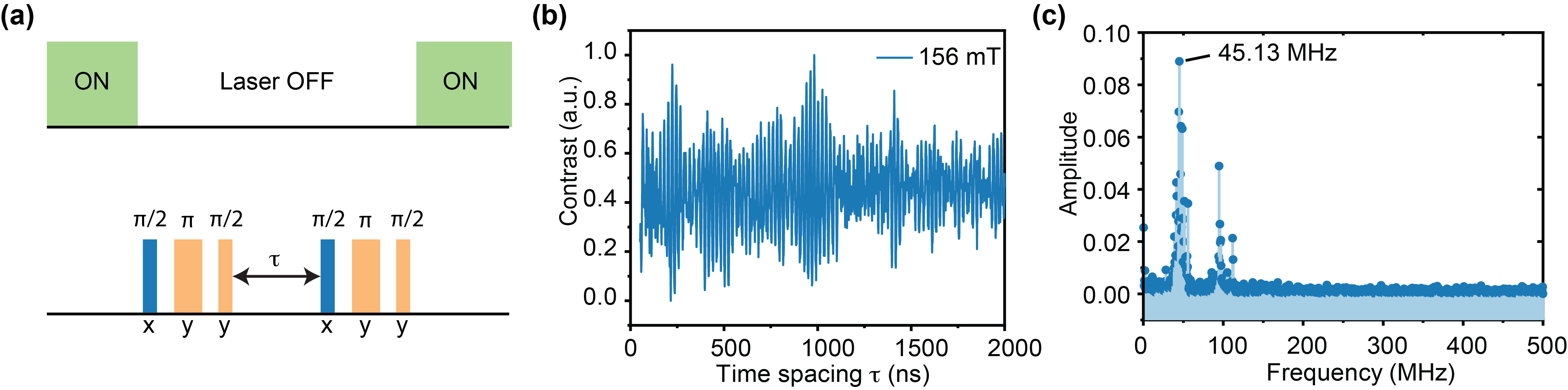}
  \caption{{(a) Sequence of correlation spectroscopy. (b) Measurement results of correlation spectroscopy and (c) its FFT results.}}
  \label{S17}    
\end{figure}

This enhanced coupling facilitated by GSLAC opens the door to optically read the state of surrounding nuclear spins. To illustrate this, we performed correlation spectroscopy to measure the coherent rotation of the surrounding nitrogen nuclear spins over periods extending beyond those typically constrained by T2. The correlation measurement sequence consists of two successive spin echo sequences and followed by optical readout, as shown in Fig. \ref{S17}(a). The time intervals between pi/2 and pi pulses within each spin echo are fixed at 28 ns, where the interaction between nuclear and electron spins is most effective in generating accumulated phase on the Bloch sphere. Time gap $\tau$ between the last pi/2 pulse of the first spin echo and the starting pi/2 pulse of the second spin echo is varied. The fluorescence is recorded as a function of $\tau$. The resulting correlation spectrum is presented in Fig. \ref{S17}(b), with its Fast Fourier Transform (FFT) analysis showcased in Fig. \ref{S17}(c), highlighting a distinct hyperfine coupling frequency of approximately 45.13 MHz.

\section{Section VIII: Supplementary Information Figures}

\begin{figure}[htpt]
  \centering
  \includegraphics[width=0.7\textwidth]{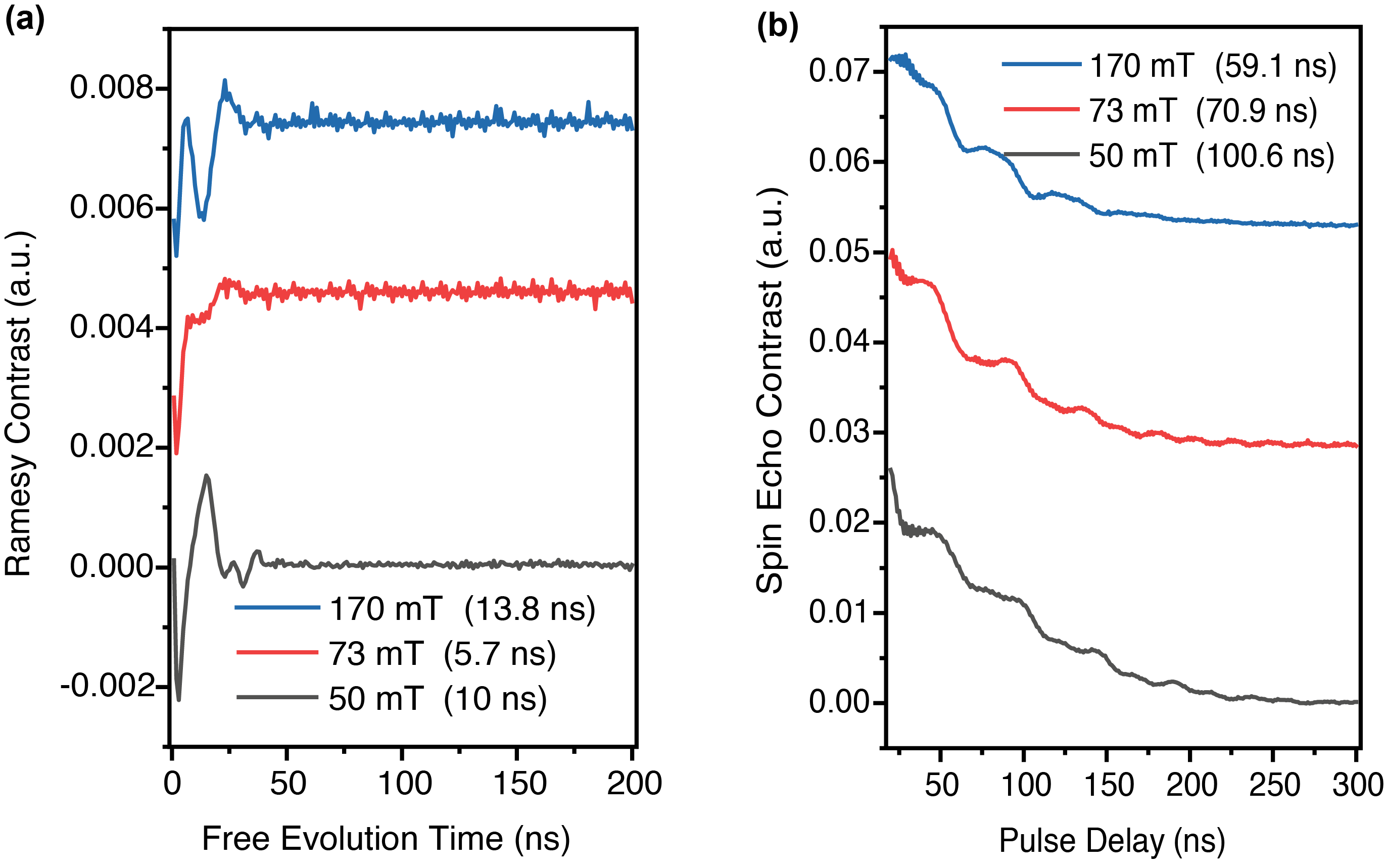}
  \caption{{(a) Ramesy and (b) Spin echo results at different magnetic fields.}}
  \label{S10}    
\end{figure}

\begin{figure}[htpt]
  \centering
  \includegraphics[width=0.5\textwidth]{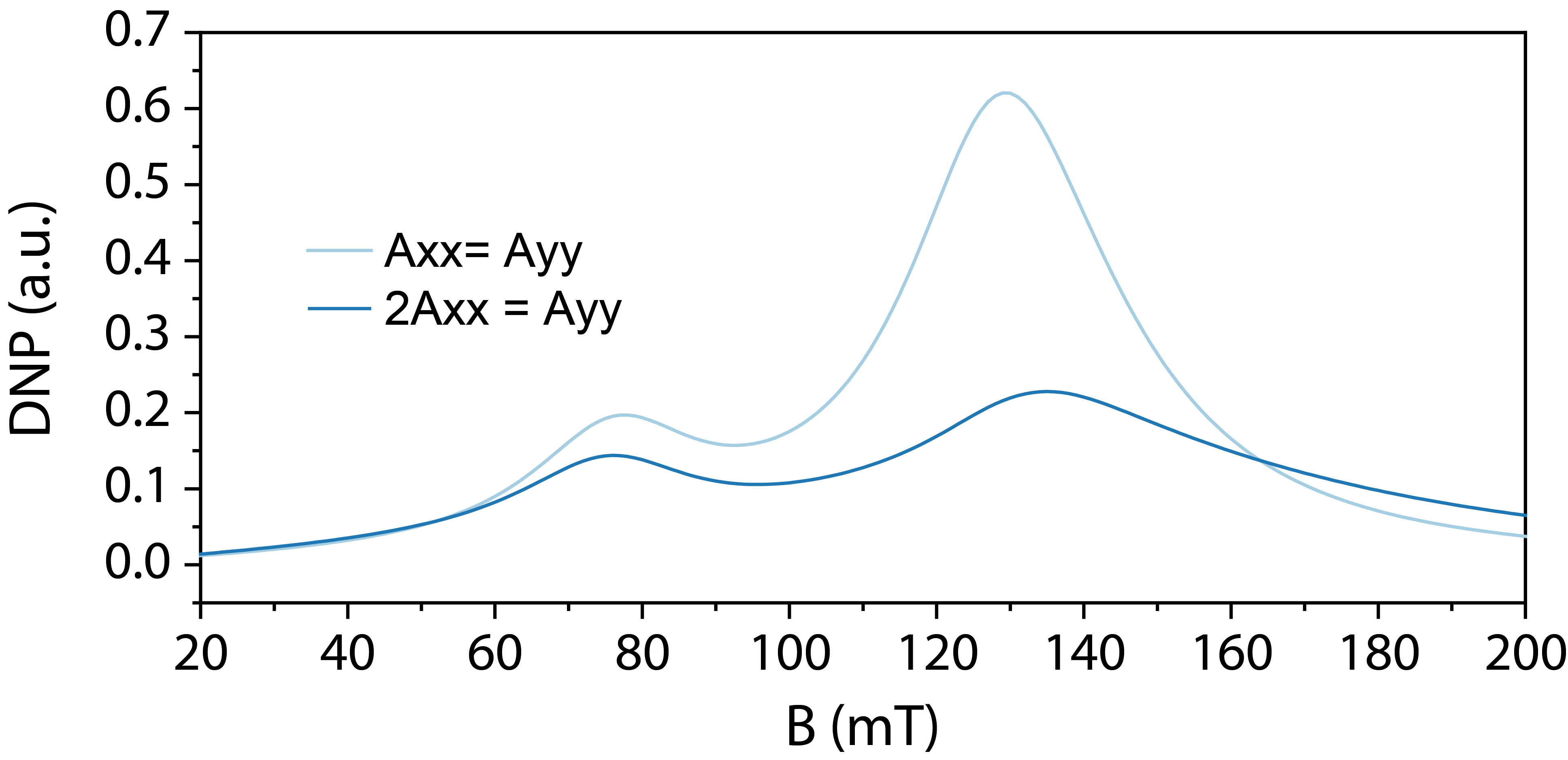}
  \caption{The simulation of nuclear polarization degree, considering the impact of the hyperfine interaction components $A_{xx}$ and $A_{yy}$.}
  \label{S13}    
\end{figure}

\begin{figure}[htpt]
  \centering
  \includegraphics[width=0.4\textwidth]{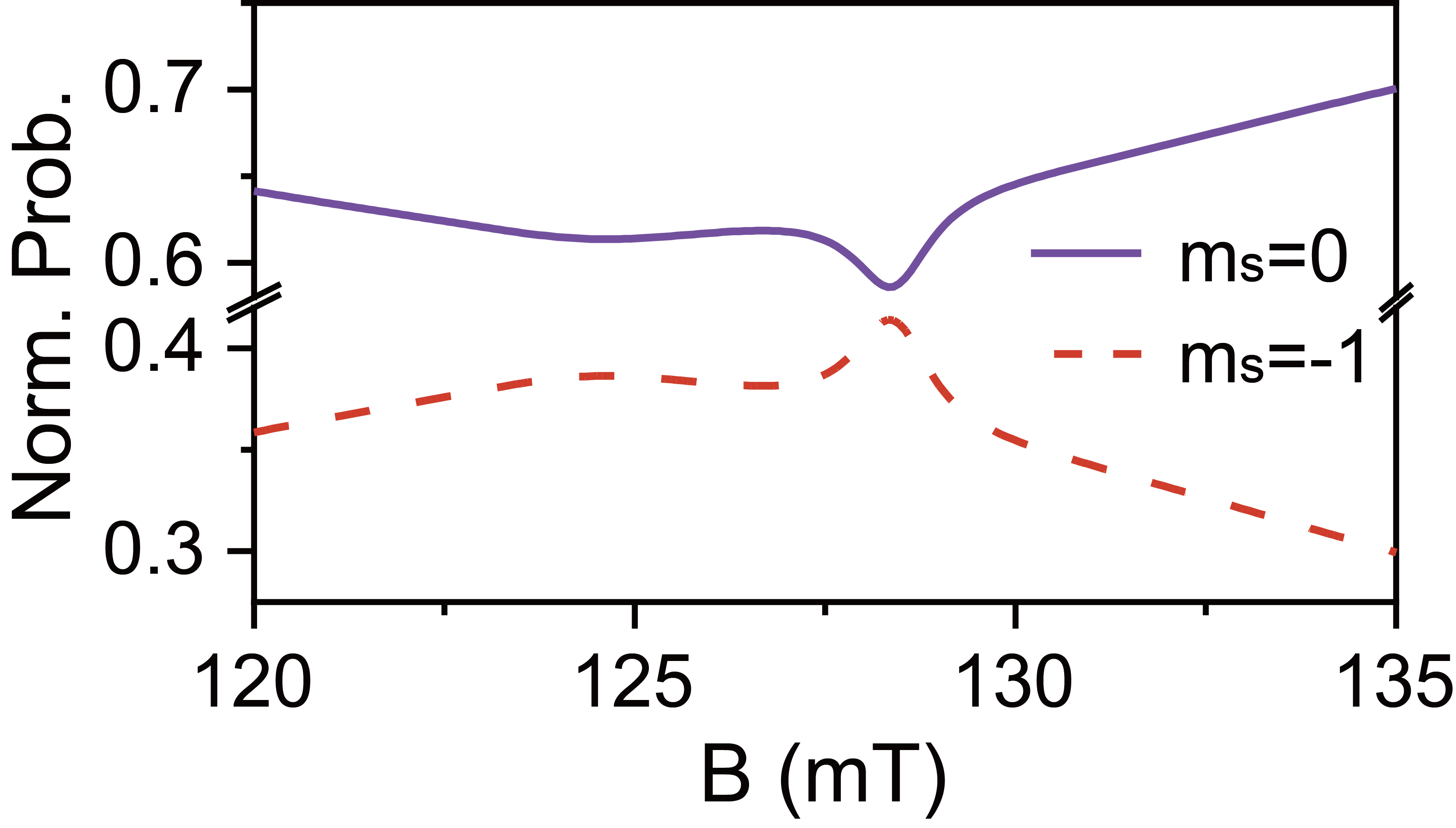}
  \caption{{The normalized probabilities of ground electronic spin states $m_s$ = 0 (purple line) and $m_s$ = -1 (red dashed line) at various magnetic field.}}
  \label{S14}    
\end{figure}

\clearpage
\newpage

\nocite{*}